\documentclass[journal]{IEEEtran}

\usepackage{amsmath,amsfonts,amssymb}
\usepackage{bm}
\usepackage[varg]{newtxmath}

\usepackage{algorithm}
\usepackage{algorithmic}

\usepackage{array}
\usepackage{booktabs}
\usepackage{multirow}
\usepackage{makecell}
\usepackage{threeparttable}
\usepackage{tabularx}
\usepackage{adjustbox}
\usepackage[table]{xcolor}

\usepackage{graphicx}
\usepackage{subcaption}
\usepackage{stfloats}
\usepackage{cuted}

\usepackage{textcomp}
\usepackage{pifont}
\usepackage{ragged2e}
\usepackage{verbatim}
\usepackage{url}
\usepackage[normalem]{ulem}
\usepackage{tikz}
\usepackage[colorlinks=true, allcolors=black, citecolor=blue]{hyperref}
\makeatletter
\renewcommand{\@cite}[2]{\textcolor{blue}{[}#1\if@tempswa , #2\fi\textcolor{blue}{]}}
\def\@citex[#1]#2{%
  \let\@citea\@empty
  \@cite{\@for\@citeb:=#2\do
    {\@citea\def\@citea{\textcolor{blue}{]}, \textcolor{blue}{[}}%
     \edef\@citeb{\expandafter\@firstofone\@citeb\@empty}%
     \if@filesw\immediate\write\@auxout{\string\citation{\@citeb}}\fi
     \@ifundefined{b@\@citeb}{\mbox{\reset@font\bfseries ?}%
       \G@refundefinedtrue
       \@latex@warning
         {Citation `\@citeb' on page \thepage \space undefined}}%
       {\hbox{\csname b@\@citeb\endcsname}}}}{#1}}
\makeatother
\usepackage{orcidlink}

\definecolor{softgray}{RGB}{242,242,242}
\definecolor{provenancegreen}{RGB}{65,145,78}
\definecolor{provenancegraygreen}{RGB}{145,162,150}
\definecolor{provenanceorange}{RGB}{226,139,43}
\definecolor{redundblue}{RGB}{0,90,220}
\definecolor{redundhl}{RGB}{191,219,254}

\providecommand{\NumSources}{\textbf{X}}

\providecommand{\NumCanonicalImages}{\textbf{X}}

\providecommand{\NumQA}{\textbf{X}}
\providecommand{\NumGrounding}{\textbf{X}}
\providecommand{\NumCaptions}{\textbf{X}}
\providecommand{\AvgImageSize}{\textbf{X}$\times$\textbf{X}}
\providecommand{\AvgCaptionLength}{\textbf{X}}

\providecommand{\NumOLUC}{\textbf{X}}
\providecommand{\NumRLUC}{\textbf{X}}
\providecommand{\NumOC}{\textbf{X}}
\providecommand{\NumRC}{\textbf{X}}
\providecommand{\NumOSR}{\textbf{X}}
\providecommand{\NumOCLS}{\textbf{X}}
\providecommand{\NumDIC}{\textbf{X}}
\providecommand{\NumFVG}{\textbf{X}}
\providecommand{\NumCBVG}{\textbf{X}}
\providecommand{\NumCCR}{\textbf{X}}

\providecommand{\MinCanonicalLongSide}{\textbf{X}}
\providecommand{\MinMosaicLongSide}{\textbf{X}}
\providecommand{\MaxHoleRatio}{\textbf{X\%}}

\providecommand{\BestZeroShot}{\textbf{X\%}}

\newcommand{\AnnHuman}{\textcolor{provenancegraygreen}{\ding{51}\kern-0.72em\ding{55}}}
\newcommand{\AnnMachine}{\textcolor{provenanceorange}{\ding{55}}}
\newcommand{\AnnSemi}{\textcolor{provenancegreen}{\ding{51}}}

\newcommand{\XLSARTableFont}{\footnotesize}
\newcommand{\XLSARTableNoteFont}{\scriptsize}
\newcommand{\XLSARTableSetup}[1]{%
  \XLSARTableFont
  \setlength{\tabcolsep}{#1}%
  \renewcommand{\arraystretch}{1.08}%
}
\newcommand{\XLSARWideTable}{\XLSARTableSetup{1.8pt}}
\newcommand{\XLSARResultTable}{\XLSARTableSetup{2.2pt}}

\newcommand{\XLSARHostedRow}{\rowcolor{softgray}}
\newcolumntype{L}[1]{>{\raggedright\arraybackslash}m{#1}}
\newcolumntype{C}[1]{>{\centering\arraybackslash}m{#1}}
\newcolumntype{Y}{>{\centering\arraybackslash}X}
\newcolumntype{H}{>{\columncolor{softgray}\centering\arraybackslash}X}

\renewcommand{\NumCanonicalImages}{405}

\renewcommand{\AvgImageSize}{7153$\times$7153}

\renewcommand{\NumQA}{14{,}250}
\renewcommand{\NumGrounding}{3260}
\renewcommand{\NumCaptions}{405}

\renewcommand{\NumOC}{747}
\renewcommand{\NumRC}{3264}
\renewcommand{\NumOLUC}{135}
\renewcommand{\NumRLUC}{395}
\renewcommand{\NumOSR}{1654}
\renewcommand{\NumOCLS}{6973}
\renewcommand{\NumDIC}{405}
\renewcommand{\NumFVG}{1943}
\renewcommand{\NumCBVG}{1317}
\renewcommand{\NumCCR}{1082}

\renewcommand{\NumSources}{11}
\renewcommand{\AvgCaptionLength}{402 / 378}
\renewcommand{\MinCanonicalLongSide}{3000}
\renewcommand{\MinMosaicLongSide}{2048}
\renewcommand{\MaxHoleRatio}{40\%}

\renewcommand{\BestZeroShot}{70.17\%}

\begin{document}

\title{XLSAR-Bench: Towards Extremely Large SAR Imagery Understanding of MLLMs}

\author{
        La Jiang$^{\orcidlink{0009-0007-6076-233X}}$, 
        Lingfeng Chen$^{\orcidlink{0009-0003-2690-9407}}$, 
        Jiaqi Zhu$^{\orcidlink{0009-0000-2232-8080}}$,
        Fengrong Guo$^{\orcidlink{0009-0000-4581-863X}}$,
        Jiehua Zhang$^{\orcidlink{0000-0003-4342-0139}}$,
        Panhe Hu$^{\orcidlink{0000-0001-5895-3152}}$, 
        Li Liu$^{\orcidlink{0000-0002-2011-2873}}$,
        and~Zhen Liu$^{\orcidlink{0000-0002-1233-1494}}$

\vspace{-2em}

\thanks{This work was supported in part by Fundamental and Interdisciplinary Disciplines Breakthrough Plan of the Ministry of Education of China under Grant JYB2025XDXM110, in part by National Natural Science Foundation of China under Grant 62401585, in part by the Science and Technology Innovation Program of Hunan Province under Grant 2024RC3120, in part by Innovation Foundation for Postgraduate of National University of Defense Technology under Grant XJJCB2026040.}
\thanks{
  Jiehua Zhang is with the College of Electronic Science and Technology, National University of Defense Technology (NUDT), Changsha 410073, China, and also with the Center for Machine Vision and Signal Analysis (CMVS), University of Oulu, 90570 Oulu, Finland (e-mail: zhangjiehua\_nudt@163.com).
  
  La Jiang, Lingfeng Chen, Jiaqi Zhu, Fengrong Guo, Panhe Hu, Li Liu and Zhen Liu are with the College of Electronic Science and Technology, National University of Defense Technology (NUDT), Changsha 410073, China (e-mail: jiangla22@nudt.edu.cn; chenlingfeng@nudt.edu.cn; zhujiaqi24@nudt.edu.cn; guofengrong21@nudt.edu.cn; hupanhe13@nudt.edu.cn;  liuli\_nudt@nudt.edu.cn; zhen\_liu@nudt.edu.cn). \textit{(La Jiang and Lingfeng Chen contributed equally to this work.)(Corresponding authors: Panhe Hu and Zhen Liu.)}}
\thanks{Digital Object Identifier XX.XXXX/XXXXX.2026.XXXXXX}
}

\markboth{SUBMITTED TO IEEE Transactions on Geoscience and Remote Sensing}%
{Jiang \MakeLowercase{\textit{\textit{et~al.}}}: XXXXXXXXXXXXXXX}

\maketitle

\begin{figure*}[!t]
  \centering
  \includegraphics[width=\textwidth]{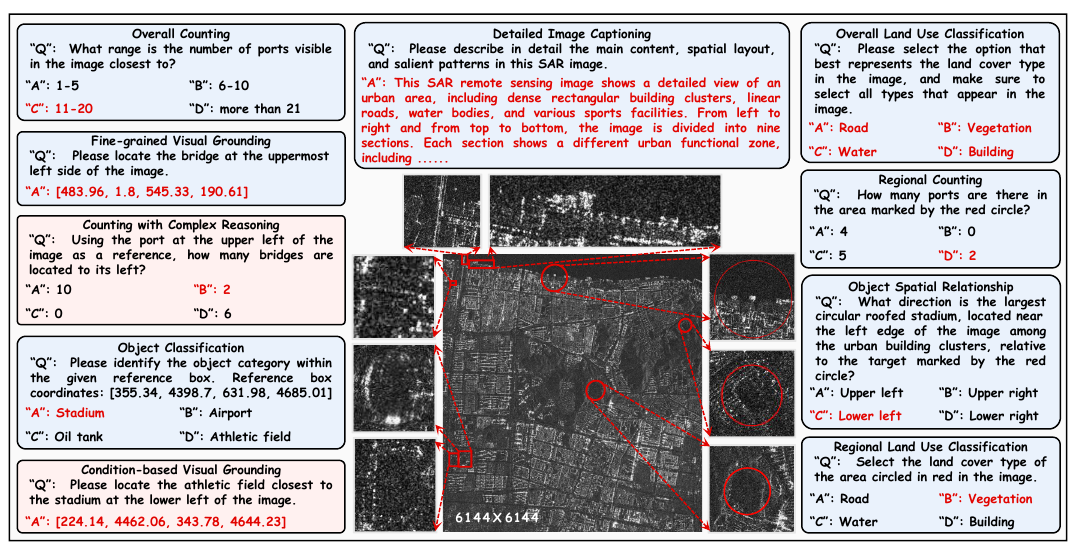}
  \caption{A representative XLSAR-Bench sample. The same large SAR image supports ten tasks involving scene interpretation, regional perception, object understanding, visual grounding, detailed description, and conditional counting.}
  \label{fig:xlsar_example}
\end{figure*}

\begin{abstract}
The rapid development of multimodal large language models (MLLMs) has highlighted the need for comprehensive benchmarks to assess their capabilities and limitations in remote sensing (RS). However, evaluating MLLMs in large-size synthetic aperture radar (SAR) scenarios remains challenging, as understanding complete scenes requires both global semantic interpretation and fine-grained target perception. Existing SAR benchmarks predominantly rely on small image patches, limiting their ability to assess these capabilities jointly. To address these limitations, we present \textit{XLSAR-Bench}, a comprehensive benchmark for evaluating the perception and reasoning capabilities of MLLMs in large-scene SAR imagery. XLSAR-Bench comprises \NumCanonicalImages{} images with an average size of \AvgImageSize{} pixels, ten tasks spanning visual question answering, visual grounding, and detailed image captioning, with all evaluation samples meticulously annotated manually, assisted by a semi-automatic captioner. Experimental results from general-purpose and domain-specific MLLMs reveal persistent challenges in understanding large SAR scenes and accurately perceiving small targets. We have open-sourced XLSAR-Bench\footnote{\href{https://huggingface.co/datasets/Jiaqikiki/XLSAR}{\texttt{\textcolor{magenta}{https://huggingface.co/datasets/Jiaqikiki/XLSAR}}}} to support further research in developing more powerful MLLMs for complete large SAR scenes.
\end{abstract}


\vspace{-0.5em}

\section{Introduction}
\label{sec:introduction}

\IEEEPARstart{M}{ultimodal} large language models (MLLMs) have demonstrated remarkable progress in visual perception and reasoning, enabling a wide range of vision-language tasks such as image captioning, visual question answering (VQA), and visual grounding. These advances have also promoted the development of MLLMs for remote sensing (RS), where models are increasingly expected to recognize objects, interpret scenes, and reason about spatial relationships through natural-language instructions~\cite{zhang2024earthgpt,zhang2025earthgptx,wang2025xlrs}. To comprehensively assess these capabilities and identify their limitations, systematic vision-language benchmarks are essential.

RS imagery plays an important role in Earth observation, supporting applications such as environmental monitoring, maritime surveillance, and urban analysis. Among different imaging modalities, SAR offers all-day and largely weather-independent observation capabilities. However, its coherent imaging mechanism and side-looking geometry introduce distinctive backscatter patterns, speckle, layover, and shadow, making visual interpretation substantially different from optical imagery~\cite{moreira2013sar_tutorial,argenti2013speckle,zhu2021deep_sar}. These challenges become more pronounced in large-scene SAR images, where small targets may occupy only a tiny fraction of the scene. Downsampling an entire image can obscure fine-grained target details, whereas cropping may remove the surrounding context necessary for scene interpretation and spatial reasoning. This raises a fundamental question: \textit{Can current MLLMs understand a complete large SAR scene while accurately perceiving its fine-grained local targets?} Despite recent progress in SAR vision-language understanding, existing benchmarks remain limited in three key aspects:

\noindent\textbf{Image Size.}
Large-scene SAR images contain extensive spatial structures and numerous small targets, requiring models to capture both global scene context and local visual details. However, existing SAR vision-language benchmarks predominantly rely on small image patches or target-centered samples~\cite{ma2025sarchat,wei2026sarlang,wang2026sareval,li2025tacmt,ma2026sarvlm,gaosarcap2025}. Although these samples facilitate object-level recognition and localized interpretation, they cannot adequately evaluate the ability of MLLMs to understand an entire large scene while retaining fine-grained target information. As shown in Table~\ref{tab:benchmark_comparison}, the compared benchmarks do not jointly assess these capabilities on complete large-scene SAR images.

\noindent\textbf{Human Annotation.}
Constructing reliable vision-language annotations for large-scene SAR imagery is challenging because of the substantial effort required to identify targets, interpret scene semantics, and verify spatial relationships. Automated annotation can improve efficiency, but generated questions and captions may contain unsupported visual attributes, ambiguous descriptions, and linguistic biases~\cite{li2023pope,parmar2023instructionbias}. Such errors can compromise the reliability of benchmark evaluation, particularly when fine-grained targets are difficult to distinguish. Therefore, large-scene SAR benchmarking requires scalable annotation procedures with explicit visual evidence and human verification.

\noindent\textbf{Evaluation Dimensions.}
Existing SAR datasets primarily focus on individual tasks, including scene classification, object detection, and semantic segmentation~\cite{ls_ssdd,sad,air_sarship,msar_bridge,whu_opt_sar,air_polsar_seg,spacenet6,fusar_map,yeseg,atrnetstar2026}. Recent SAR vision-language resources have extended evaluation to captioning, question answering, dialogue, and grounding. Nevertheless, their task coverage remains fragmented, and the joint assessment of scene-level understanding, regional perception, object counting, spatial relationships, and fine-grained localization within the same large image is still limited. A comprehensive benchmark is therefore needed to evaluate these complementary capabilities under a unified setting.

\begin{figure}[t]
    \centering
    \includegraphics[width=0.94\columnwidth]{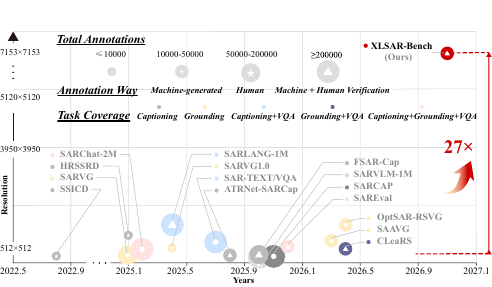}
    \caption{XLSAR-Bench surpasses existing SAR vision-language datasets in terms of image extent and task coverage. The 27$\times$ ratio is the XLSAR-Bench average long side (7{,}153 pixels) over the count-weighted average image size of existing SAR vision-language datasets (258.9 pixels).}
    \label{fig:benchmark_positioning}
\end{figure}

To address these challenges, we introduce \textit{XLSAR-Bench}, a benchmark for evaluating the perception and reasoning capabilities of MLLMs in large-scene SAR imagery. We collect \NumCanonicalImages{} images from \NumSources{} SAR data sources, with an average size of \AvgImageSize{} pixels. Based on source bounding boxes, masks, and polygons, we construct ten tasks in three vision-language formats: VQA, visual grounding, and detailed image captioning. The annotations are further verified by human annotators. We evaluate general-purpose, RS-oriented, and SAR-specific MLLMs and establish \textit{XLSAR-Baseline}, a simple whole-image LoRA adaptation baseline. The key advantages of XLSAR-Bench are threefold: \textit{i) Large-scene imagery.} Both global scene understanding and fine-grained target perception are evaluated within complete large SAR images; \textit{ii) High-quality annotation.} All annotations derive from existing spatial labels and are verified against the original SAR imagery; captions are generated with a semi-automatic captioner and meticulous manual annotation; and \textit{iii) Comprehensive evaluation dimensions.} Ten tasks cover scene interpretation, object recognition, counting, spatial reasoning, grounding, and detailed captioning.

In summary, our main contributions are as follows:
\begin{itemize}
    \item We establish XLSAR-Bench, a new benchmark for evaluating MLLMs across ten tasks in large-size SAR scenarios.

    \item We develop a label-driven annotation pipeline with meticulous manual annotation and a semi-automatic captioner to construct reliable vision-language evaluation samples.

    \item We evaluate representative MLLMs and provide a simple LoRA adaptation baseline, revealing their capabilities and limitations in large-scene SAR understanding.
\end{itemize}

\section{Related Work}
\label{sec:related_work}

\noindent\textbf{General Multimodal Benchmark.}
Large-scale vision-language models (VLMs) have demonstrated remarkable capabilities in visual perception and reasoning, promoting the development of comprehensive multimodal benchmarks. Early evaluations mainly focused on individual tasks (\textit{e.g.}, image captioning and VQA), while recent benchmarks have expanded to more diverse capabilities. MME evaluates multiple perception and cognition dimensions, and MMBench provides a systematic assessment of fine-grained multimodal abilities. Seed-Bench further expands the evaluation scale, while MMT-Bench incorporates diverse real-world scenarios. More recently, MME-RealWorld extends multimodal evaluation to high-resolution natural images with an average size of approximately 2,000$\times$1,500 pixels~\cite{zhang2024mmerealworld}. Nevertheless, these benchmarks primarily target natural imagery and provide limited evaluation of SAR-specific visual understanding. Their image sizes are also considerably smaller than those encountered in large-size remote-sensing applications.

\noindent\textbf{Remote Sensing Multimodal Benchmark.}
With the rapid development of MLLMs, RS has witnessed increasing interest in vision-language understanding. Early datasets primarily addressed specific tasks (\textit{e.g.}, image captioning with RSICD~\cite{lu2018rsicd}, VQA with RSVQA~\cite{lobry2020rsvqa}, and visual grounding with RSVG~\cite{zhan2023rsvg}). More recent benchmarks provide broader evaluation dimensions. VRSBench~\cite{li2024vrsbench} combines image descriptions, object references, and question-answer pairs, while XLRS-Bench~\cite{wang2025xlrs} introduces 16 sub-tasks for evaluating perception and reasoning in extremely large RS images. In the SAR domain, conventional datasets mainly support individual interpretation tasks (\textit{e.g.}, ship detection with LS-SSDD~\cite{ls_ssdd} and AIR-SARShip~\cite{air_sarship}, bridge detection with MSAR-Bridge~\cite{msar_bridge}, and semantic segmentation with WHU-OPT-SAR~\cite{whu_opt_sar} and SpaceNet6~\cite{spacenet6}). Recent efforts such as ATRNet-STAR~\cite{atrnetstar2026} and SARATR-X~\cite{saratrx2025} further advance fine-grained recognition and transferable SAR representations. Meanwhile, SAR-specific vision-language resources have expanded toward captioning, dialogue, grounding, and question answering (\textit{e.g.}, SARChat~\cite{ma2025sarchat}, SARLANG~\cite{wei2026sarlang}, and SAR-oriented evaluation benchmarks~\cite{wang2026sareval,weng2026clears}), alongside broader SAR-text supervision~\cite{gaosarcap2025,he2025sartext,zhang2026fsarcap,jiang2025sarclip,ma2026sarvlm}. Nevertheless, these resources have two main limitations. First, their evaluation tasks are often separated across datasets, limiting comprehensive assessments of multiple capabilities. Second, most SAR vision-language evaluation samples remain patch-based or target-centered, making it difficult to assess global scene understanding and fine-grained target perception within complete large-scene images. Table~\ref{tab:benchmark_comparison} summarizes representative SAR vision-language resources, and Figure~\ref{fig:benchmark_positioning} contrasts XLSAR-Bench with them in terms of image extent and task coverage.

\begin{table*}[!t]
\centering
\caption{Comparison of SAR vision--language resources and XLSAR-Bench. \AnnHuman{}, \AnnMachine{}, and \AnnSemi{} denote human, machine-generated, and machine-generated then human-verified annotations.}
\label{tab:benchmark_comparison}
\begin{threeparttable}
\XLSARTableFont
\setlength{\tabcolsep}{1.2pt}
\renewcommand{\arraystretch}{1.02}
\begin{tabular}{@{}L{85pt}|C{80pt}|C{31pt}|C{49pt}C{50pt}C{38pt}|C{46pt}C{31pt}|C{52pt}C{\dimexpr\textwidth-483.6pt\relax}@{}}
\toprule
Dataset & Resolution (px) & Lang. & \# Images &
\multicolumn{2}{c}{Detailed caption} &
\multicolumn{2}{c}{Visual grounding} &
\multicolumn{2}{c}{VQA} \\
\cmidrule(lr){5-6}\cmidrule(lr){7-8}\cmidrule(lr){9-10}
& & & & \# Texts & Avg. words & \# Expr. & Ann. & \# QA & Ann. \\
\midrule
SSICD~\cite{ssicd2022} & $256^2$; var. & EN & 1,500 & 7,500 & -- & -- & -- & -- & -- \\
ATRNet-SARCap~\cite{gaosarcap2025} & $224^2$ & EN & 5,251 & 47,259 & 11.6 & -- & -- & -- & -- \\
HRSSRD~\cite{li2025hrssrd} & $1{,}024^2$ & EN & 1,000 & 5,000 & -- & -- & -- & -- & -- \\
SARVLM-1M~\cite{ma2026sarvlm} & $224^2$ (eval) & EN & 365,984 & 1,727,408 & -- & -- & -- & -- & -- \\
SARCAP~\cite{jiang2025sarclip} & $\leq1{,}024^2 / 224^2$ & EN & -- & $>400{,}000$ & -- & -- & -- & -- & -- \\
FSAR-Cap~\cite{zhang2026fsarcap} & $1{,}024^2 / 224^2$ & EN & 14,480 & 72,400 & 20.8 & -- & -- & -- & -- \\
SARVG1.0~\cite{li2025tacmt} & $512^2$ & EN & 2,465 & -- & -- & 7,617 & \AnnHuman & -- & -- \\
SARVG~\cite{zhan2025vgrss} & $256^2$ & EN & 43,798 & -- & -- & 54,429 & \AnnMachine & -- & -- \\
SAAVG~\cite{bui2026saavg} & $800^2$ & EN & 10,001 & -- & -- & 20,652 & \AnnSemi & -- & -- \\
OptSAR-RSVG~\cite{tang2026optsar} & $640^2$ & EN & 12,466 & -- & -- & 20,178 & \AnnSemi & -- & -- \\
SARChat-2M~\cite{ma2025sarchat} & $448^2$ & EN & -- & 52,173 & -- & 107,189 & \AnnMachine & 1,904,186 & \AnnMachine \\
SAR-TEXT/VQA~\cite{he2025sartext} & $256^2/512^2/800^2$ & EN & 136,584 & 136,584 & 16.3 & -- & -- & $>130$k & \AnnMachine \\
SARLANG-1M~\cite{wei2026sarlang} & $\leq1{,}024^2$ & EN & 118,331 & 45,650 & -- & -- & -- & 1,080,627 & \AnnSemi \\
SAREval~\cite{wang2026sareval} & $512^2$ & EN & -- & 395 & -- & 3,605 & \AnnSemi & 6,925 & \AnnSemi \\
CLeaRS~\cite{weng2026clears} & $504^2$ & EN & 10,157/10,000 & -- & -- & 12,706 & \AnnSemi & 25,000 & \AnnSemi \\
\midrule
\textbf{XLSAR-Bench (ours)} & $\sim2$k--$40$k (long-side) & EN / ZH & \NumCanonicalImages & \NumCaptions & \AvgCaptionLength & \NumGrounding & \AnnSemi & \NumQA & \AnnSemi \\
\bottomrule
\end{tabular}
\begin{tablenotes}[flushleft]
\XLSARTableNoteFont
\item[] Resolution lists released/source sizes; explicit evaluation input follows the slash for SARCAP and FSAR-Cap, while ``eval'' marks SARVLM. SSICD combines fixed and variable source images. CLeaRS count slashes denote VG/VQA; XLSAR is an audited long-side source envelope; a dash marks unavailable or inapplicable fields.
\end{tablenotes}
\end{threeparttable}
\end{table*}

\noindent\textbf{Multimodal Large Language Model.}
Recent advances in MLLMs have significantly improved multimodal perception and reasoning. General-purpose models such as Qwen2-VL~\cite{wang2024qwen2vl}, Qwen2.5-VL~\cite{bai2025qwen25vl}, and InternVL~\cite{chen2024internvl15} demonstrate strong visual understanding across diverse tasks. To address the limitations of input resolution, several approaches have been developed for high-resolution image processing. For example, LLaVA-UHD~\cite{guo2024llavauhd} divides high-resolution images into smaller patches and compresses visual features, while TokenPacker~\cite{li2025tokenpacker} reduces visual-token redundancy. Other approaches leverage text-guided token selection or adaptive cropping to preserve relevant image details~\cite{luo2025largeRSI,liu2026zoomearth}. In the RS field, GeoChat~\cite{kuckreja2024geochat} enables grounded dialogue, LHRS-Bot~\cite{muhtar2024lhrsbot} enhances multimodal instruction following, and EarthGPT~\cite{zhang2024earthgpt} and EarthGPT-X~\cite{zhang2025earthgptx} extend understanding across remote-sensing modalities. Parameter-efficient adaptation methods, such as LoRA~\cite{hu2022lora,luo2023lavin}, further facilitate domain-specific model adaptation. Despite these advancements, general-purpose and RS-oriented MLLMs have not been systematically evaluated on complete large-size SAR scenes across both global and fine-grained language tasks.

\noindent\textbf{Our Insight - }Existing SAR benchmarks primarily evaluate localized visual understanding, while high-resolution MLLMs focus on preserving image details through specialized processing strategies. However, their ability to jointly understand complete SAR scenes and fine-grained local targets remains insufficiently explored. In contrast, XLSAR-Bench provides a unified benchmark spanning VQA, visual grounding, and detailed image captioning on large-scene SAR imagery. We further establish a simple whole-image LoRA baseline to investigate the effectiveness of standard adaptation without specialized high-resolution architectures.


\setcounter{topnumber}{2}
\setcounter{bottomnumber}{1}
\setcounter{totalnumber}{3}

\section{XLSAR-Bench}
\label{sec:xlsar_bench}

XLSAR-Bench is designed to evaluate the perception and reasoning capabilities of MLLMs in large-size SAR scenarios. It features four key characteristics: \textit{i)} large-scene SAR images that preserve global scene context and fine-grained targets; \textit{ii)} ten tasks covering diverse perception and reasoning capabilities; \textit{iii)} label-supported annotations with human verification; and \textit{iv)} bilingual evaluation in English and Chinese. Further details are presented below.

\subsection{Evaluation Dimensions}
\label{subsec:task_taxonomy}

\begin{figure}[!t]
    \centering
    \includegraphics[width=0.9\columnwidth]{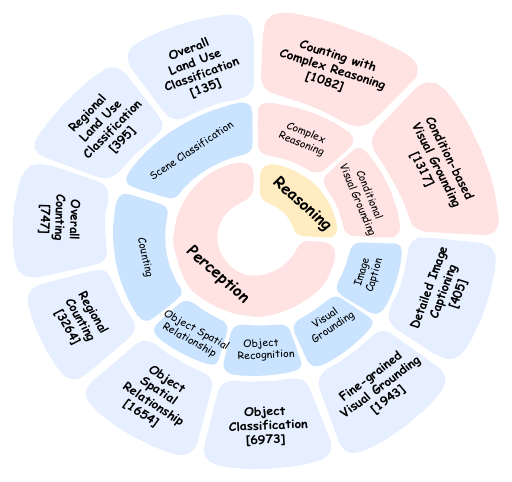}
    \caption{Task organization of XLSAR-Bench, comprising ten perception and reasoning tasks across VQA, visual grounding, and detailed image captioning.}
    \label{fig:task_taxonomy}
\end{figure}

Understanding large-scene SAR imagery requires MLLMs to recognize both global scene structures and fine-grained local targets. Following XLRS-Bench~\cite{wang2025xlrs}, we organize the evaluation into two primary capabilities: perception and reasoning. These are further divided into ten sub-tasks, comprising seven VQA tasks, two visual grounding tasks, and one detailed captioning task. Figure~\ref{fig:task_taxonomy} illustrates the task organization, with detailed definitions summarized in Table~\ref{tab:task_taxonomy}.

\subsubsection{Perception}

We employ image captioning, VQA, and visual grounding to evaluate the perception capabilities of MLLMs at scene, region, and object levels. Specifically, eight sub-tasks are defined across six dimensions:

\begin{itemize}
    \item \textbf{Image Captioning.} Detailed Image Captioning (DIC) requires comprehensive descriptions of the entire SAR scene, including its overall structure, regional characteristics, and local object details.

    \item \textbf{Scene Classification.} This dimension includes Overall Land Use Classification (OLUC) and Regional Land Use Classification (RLUC). The former identifies land-use categories across the whole image, while the latter focuses on specified regions.

    \item \textbf{Counting.} Overall Counting (OC) evaluates object counting across the complete image, whereas Regional Counting (RC) restricts counting to a designated area.

    \item \textbf{Object Classification.} Object Classification (OCLS) requires identifying the category of a specified target, particularly small objects embedded in large SAR scenes.

    \item \textbf{Object Spatial Relationship.} This task (OSR) evaluates the ability to identify objects and determine their relative spatial positions.

    \item \textbf{Visual Grounding.} Fine-grained Visual Grounding (FVG) requires locating a target described in natural language, with particular emphasis on small targets in large images.
\end{itemize}

\subsubsection{Reasoning}

Beyond direct visual perception, we evaluate the ability of MLLMs to combine multiple spatial constraints and infer target locations or counts. Two sub-tasks are designed:

\begin{itemize}
    \item \textbf{Complex Reasoning.} Counting with Complex Reasoning (CCR) requires counting objects that satisfy multiple conditions (\textit{e.g.}, a specified object category within a particular region).

    \item \textbf{Visual Grounding.} Condition-based Visual Grounding (CBVG) requires locating a target according to combined semantic and spatial conditions (\textit{e.g.}, an object satisfying a specified spatial relationship).
\end{itemize}

\begin{table*}[!t]
\centering
\caption{L3 task taxonomy of XLSAR-Bench. Counts are language-independent semantic instances with matched EN/ZH versions.}
\label{tab:task_taxonomy}
\XLSARWideTable
\begin{tabular*}{\textwidth}{@{\extracolsep{\fill}}
  L{0.070\textwidth}|
  L{0.120\textwidth}|
  L{0.235\textwidth}
  C{0.050\textwidth}
  C{0.100\textwidth}
  C{0.105\textwidth}
  C{0.075\textwidth}
  L{0.140\textwidth}@{}}
\toprule
\multicolumn{1}{c}{L1-Task} & \multicolumn{1}{c}{L2-Task} &
\multicolumn{1}{c}{L3-Task} & Abbr. &
\makecell[c]{Annotation\\Format} & \makecell[c]{Annotation\\Method} &
\makecell[c]{Number of\\Samples} & \multicolumn{1}{c}{Answer Type} \\
\midrule
\multirow{8}{*}{Perception}
& \multirow{2}{*}{Counting}
& Overall Counting & OC & VQA & Semi-automated & \NumOC & Multiple choice \\
& & Regional Counting & RC & VQA & Semi-automated & \NumRC & Multiple choice \\
\cmidrule(lr){2-8}
& \multirow{2}{*}{\makecell[l]{Scene\\classification}}
& Overall Land Use Classification & OLUC & VQA & Semi-automated & \NumOLUC & Multiple choice \\
& & Regional Land Use Classification & RLUC & VQA & Semi-automated & \NumRLUC & Multiple choice \\
\cmidrule(lr){2-8}
& \makecell[l]{Object spatial\\relationship}
& Object Spatial Relationship & OSR & VQA & Semi-automated & \NumOSR & Multiple choice \\
\cmidrule(lr){2-8}
& Object properties
& Object Classification & OCLS & VQA & Semi-automated & \NumOCLS & Multiple choice \\
\cmidrule(lr){2-8}
& Image captioning
& Detailed Image Captioning & DIC & Caption & Semi-automated & \NumDIC & Plain text \\
\cmidrule(lr){2-8}
& \makecell[l]{Fine-grained\\grounding}
& Fine-grained Visual Grounding & FVG & Visual grounding & Semi-automated & \NumFVG & Bounding box \\
\midrule
\multirow{2}{*}{Reasoning}
& \makecell[l]{Condition-based\\grounding}
& Condition-based Visual Grounding & CBVG & Visual grounding & Semi-automated & \NumCBVG & Bounding box \\
\cmidrule(lr){2-8}
& Complex reasoning
& Counting with Complex Reasoning & CCR & VQA & Semi-automated & \NumCCR & Multiple choice \\
\bottomrule
\end{tabular*}
\end{table*}

\subsection{Data Collection and Quality Control}
\label{subsec:source_canonicalization}

\noindent\textbf{Data Source.}
To construct a benchmark with diverse SAR scenes and evaluation tasks, we collect images from \NumSources{} public SAR datasets, including four detection and seven segmentation sources. Detection datasets (\textit{e.g.}, LS-SSDD~\cite{ls_ssdd}, HRSID~\cite{hrsid}, AIR-SARShip~\cite{air_sarship}, and MSAR-Bridge~\cite{msar_bridge}) provide object-level annotations. Segmentation datasets (\textit{e.g.}, WHU-OPT-SAR~\cite{whu_opt_sar}, AIR-PolSAR-Seg~\cite{air_polsar_seg}, SpaceNet6~\cite{spacenet6}, FUSAR-Map~\cite{fusar_map}, EarthMiss~\cite{earthmiss}, YESeg-OPT-SAR~\cite{yeseg}, and AIR-MDSAR-MAP~\cite{air_mdsar_map}) provide regional semantic information and, where available, building polygons. After mosaicking and deduplication, the benchmark retains \NumCanonicalImages{} unique large-scene images and 17{,}915 task instances. Their sizes span 174 distinct dimensions, with long sides ranging from 2{,}048 to 39{,}936 pixels; the largest single image (EarthMiss) reaches 39{,}936$\times$22{,}022, approximately $8.8\times10^{8}$ pixels. The 15 LS-SSDD-v1.0 Sentinel-1 scenes arrive as complete scenes of 24{,}000$\times$16{,}000, the 100 WHU-OPT-SAR scenes at 5{,}556$\times$3{,}704, the 31 AIR-SARShip-1.0 images at 3{,}000$^2$ to 4{,}140$^2$, and the 84 MSAR-1.0 scenes mostly at 3{,}072$^2$ to 5{,}120$^2$. Among the segmentation sources, EarthMiss contributes 13 scenes of up to 39{,}936 pixels wide, FUSAR-Map 13 scenes spanning 2{,}048 to 15{,}360 pixels, SpaceNet 6 5 strips of 3{,}100 to 9{,}700 pixels wide, AIR-MDSAR-Map (Ka band) 4 scenes up to 24{,}582$\times$8{,}522, AIR-PolSAR-Seg-2.0 3 scenes up to 7{,}820$\times$6{,}488, and YESeg-OPT-SAR 3 scenes up to 4{,}096$\times$2{,}048, with HRSID supplying the remaining 134 detection scenes up to 21{,}500 pixels. To the best of our knowledge, this is the first SAR vision-language evaluation conducted on complete scenes at this scale.

\begin{figure}[!t]
    \centering
    \includegraphics[width=\columnwidth]{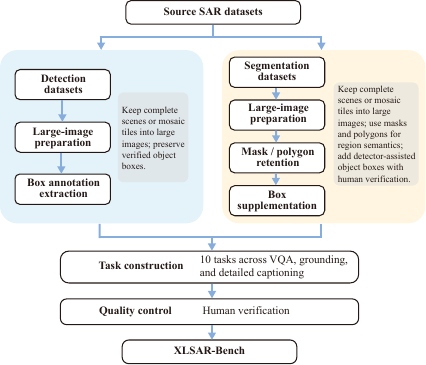}
    \caption{Construction pipeline of XLSAR-Bench: image preparation, annotation generation, and quality control.}
    \label{fig:benchmark_pipeline}
\end{figure}

To preserve large-scene information, we select complete images or reconstruct original scenes from spatially aligned tiles. Directly selected images must have a long side of at least \MinCanonicalLongSide{} pixels. Reconstructed scenes must satisfy the minimum size of \MinMosaicLongSide{} pixels and an unobserved-area ratio below \MaxHoleRatio{}. For segmentation-only scenes, additional object boxes are obtained through detector-assisted annotation and human verification. Only SAR imagery is presented to evaluated models, while paired optical data, when available, are used solely for annotation verification. Figure~\ref{fig:benchmark_pipeline} illustrates the construction process.

\noindent\textbf{Visual Question Answering.}
XLSAR-Bench provides seven VQA sub-tasks in multiple-choice format. Questions and reference answers are derived from source annotations, including object categories, counts, land-use labels, and spatial relationships. For instance, counting questions are generated from verified object instances, while regional classification questions use semantic masks. Distractors are constructed according to the answer type and checked to ensure answer uniqueness.

\noindent\textbf{Visual Grounding.}
We evaluate visual grounding through FVG and CBVG. Each instance consists of a referring expression and its corresponding target box in the original image coordinate system. FVG describes identifiable objects using their categories and visual or spatial characteristics, while CBVG introduces additional conditions to distinguish the target from similar objects. Ambiguous references are excluded through verification.

\noindent\textbf{Image Captioning.}
Unlike conventional short captions, large-size SAR scenes require detailed descriptions of global structures and local characteristics. We therefore adopt a general-to-specific format for DIC, describing the overall scene, regional layouts, and important objects. Caption drafts are generated by a semi-automatic captioner and refined through meticulous manual annotation. In particular, object categories and counts are checked against accepted annotations, while visible SAR patterns and spatial arrangements are verified against the imagery. Unsupported semantic interpretations are removed.

\noindent\textbf{Quality Control.}
All task instances undergo automated consistency checks for annotation geometry, class mapping, answer correctness, and target references. Human verification further examines semantic accuracy and visual evidence: a second reviewer re-examines 700 instances, yielding an agreement of 83.1\%, and unresolved cases are adjudicated by 3 senior reviewers. English and Chinese annotations share the same reference answers and target geometries. To reduce information leakage, images and task instances originating from the same source scene are assigned to the same data partition.

\subsection{Analysis}
\label{subsec:split_protocol}

\noindent\textbf{Image Size.}
In practice, SAR scenes typically cover city-level areas or above, with long sides often exceeding 10{,}000 pixels. However, existing SAR vision-language benchmarks evaluate MLLMs on cross-modal perception and understanding (VQA, DIC, and visual grounding) using small image slices, typically 256$\times$256 to 800$\times$800 pixels, as summarized in Table~\ref{tab:benchmark_comparison}. Such slices fail to assess the models' capability to capture long-range spatial-semantic relationships across distant regions. In contrast, XLSAR-Bench contains \NumCanonicalImages{} large-scene SAR images with average dimensions of \AvgImageSize{} pixels, preserving extensive scene structures and small targets within the same evaluation unit and enabling the joint assessment of global and local understanding.

\noindent\textbf{Dataset Composition.}
XLSAR-Bench incorporates \NumQA{} VQA instances, \NumGrounding{} visual grounding instances, and \NumCaptions{} detailed captions across ten sub-tasks. The benchmark includes both detection-oriented and segmentation-oriented scenes, providing complementary object-level and region-level annotations. The task distributions are reported in Table~\ref{tab:task_taxonomy}.

\noindent\textbf{Length of Caption.}
XLSAR-Bench provides detailed descriptions averaging 402 and 378 words in English and Chinese, respectively. Each caption follows a general-to-specific structure, covering the overall scene, regional characteristics, and identifiable local objects.

\noindent\textbf{Bilingual Evaluation.}
XLSAR-Bench supports English and Chinese evaluation with consistent image content, reference answers, and target annotations. Both language versions follow the same source-scene-level data partitions, enabling comparisons under a unified evaluation setting.

\setcounter{topnumber}{3}
\setcounter{bottomnumber}{2}
\setcounter{totalnumber}{5}
\setcounter{dbltopnumber}{2}
\renewcommand{\dbltopfraction}{0.92}
\renewcommand{\topfraction}{0.95}

\begin{table*}[!t]
\centering
\caption{Zero-shot L2 VQA results on XLSAR-Bench in EN/ZH. Scores and pooled Avg. are reported in percent. Proprietary models are highlighted in gray. The best score is in \textbf{bold} and the second-best is \underline{underlined}.}
\label{tab:zeroshot_vqa}
\XLSARResultTable
\begin{tabularx}{\textwidth}{@{}L{0.155\textwidth}L{0.135\textwidth}C{0.070\textwidth}|YYYY|Y|Y@{}}
\toprule
\multirow{2}{*}{Method} &
\multirow{2}{*}{LLM} &
\multirow{2}{*}{Lang.} &
\multicolumn{4}{c}{Perception} &
Reasoning &
\multirow{2}{*}{Avg.} \\
\cmidrule(lr){4-7}\cmidrule(lr){8-8}
& & &
Counting &
\makecell{Scene\\classification} &
\makecell{Object spatial\\relationship} &
\makecell{Object\\properties} &
\makecell{Complex\\reasoning} &
\\
\midrule
\XLSARHostedRow GLM-5.3-flash & -- & EN & \textbf{51.83} & \underline{40.00} & \textbf{59.49} & \textbf{84.76} & \textbf{39.46} & \textbf{67.45} \\
\XLSARHostedRow GPT-6.1-sol & -- & EN & \underline{48.19} & \textbf{45.85} & \underline{47.70} & \underline{80.82} & \underline{38.26} & \underline{63.26} \\
Qwen2.5-VL & Qwen2.5-7B & EN & 41.88 & 39.81 & 32.53 & 57.44 & 28.37 & 47.31 \\
LLaVA-OV & Qwen2-7B & EN & 40.17 & 30.94 & 30.41 & 58.61 & 29.48 & 46.90 \\
LLaVA-1.5 & Vicuna-v1.5-7B & EN & 28.65 & 20.38 & 28.78 & 22.56 & 24.12 & 25.03 \\
InternVL3 & Qwen2.5-7B & EN & 46.30 & 32.26 & 19.56 & 64.76 & 30.39 & 50.50 \\
GeoChat & Vicuna-v1.5-7B & EN & 28.15 & 23.40 & 34.02 & 21.91 & 34.01 & 26.05 \\
SARChat & GLM-Edge-4B-Chat & EN & 45.34 & 24.82 & 43.83 & 80.21 & 36.14 & 60.77 \\
\midrule
\XLSARHostedRow GLM-5.3-flash & -- & ZH & \textbf{55.58} & 40.02 & \textbf{63.25} & \textbf{86.56} & \textbf{43.97} & \textbf{70.17} \\
\XLSARHostedRow GPT-6.1-sol & -- & ZH & 47.79 & \textbf{42.27} & \underline{48.07} & \underline{82.70} & \underline{39.83} & \underline{64.10} \\
Qwen2.5-VL & Qwen2.5-7B & ZH & 42.76 & \underline{41.43} & 36.78 & 70.79 & 30.87 & 54.83 \\
LLaVA-OV & Qwen2-7B & ZH & 44.98 & 31.70 & 29.63 & 71.68 & 29.11 & 54.56 \\
LLaVA-1.5 & Vicuna-v1.5-7B & ZH & 28.65 & 20.20 & 28.78 & 21.71 & 24.12 & 24.61 \\
InternVL3 & Qwen2.5-7B & ZH & \underline{47.88} & 28.94 & 20.61 & 72.48 & 30.87 & 54.76 \\
GeoChat & Vicuna-v1.5-7B & ZH & 28.50 & 20.01 & 28.90 & 21.71 & 24.12 & 24.57 \\
SARChat & GLM-Edge-4B-Chat & ZH & 46.47 & 26.60 & 47.64 & 77.01 & 37.34 & 60.12 \\
\bottomrule
\end{tabularx}
\end{table*}

\begin{table*}[!t]
\centering
\caption{Zero-shot detailed-captioning results on XLSAR-Bench in EN/ZH. Scores are reported in percent. Proprietary models are highlighted in gray. The best score is in \textbf{bold} and the second-best is \underline{underlined}.}
\label{tab:zeroshot_caption}
\XLSARResultTable
\begin{tabularx}{\textwidth}{@{}L{0.155\textwidth}L{0.135\textwidth}C{0.070\textwidth}|YYYYYY@{}}
\toprule
Method & LLM & Lang. & BLEU-1 & BLEU-2 & BLEU-3 & BLEU-4 & METEOR & ROUGE-L \\
\midrule
\XLSARHostedRow GLM-5.3-flash & -- & EN & 19.18 & 9.40 & 4.14 & 1.75 & \textbf{20.46} & 15.09 \\
\XLSARHostedRow GPT-6.1-sol & -- & EN & 17.89 & 8.70 & 3.28 & 1.42 & \underline{18.26} & 17.19 \\
Qwen2.5-VL & Qwen2.5-7B & EN & 20.69 & 8.42 & 3.42 & 1.40 & 17.84 & 15.29 \\
LLaVA-OV & Qwen2-7B & EN & \underline{29.97} & 12.19 & 5.07 & \underline{2.04} & 14.00 & 16.30 \\
LLaVA-1.5 & Vicuna-v1.5-7B & EN & \textbf{29.98} & \textbf{12.87} & \underline{5.09} & 1.86 & 12.11 & \textbf{18.05} \\
InternVL3 & Qwen2.5-7B & EN & 26.20 & 10.72 & 4.52 & 1.88 & 15.13 & 14.62 \\
GeoChat & Vicuna-v1.5-7B & EN & 28.72 & \underline{12.54} & \textbf{5.25} & \textbf{2.13} & 12.83 & \underline{17.54} \\
\midrule
\XLSARHostedRow GLM-5.3-flash & -- & ZH & 17.33 & 6.69 & 2.53 & 0.97 & \textbf{24.04} & 12.60 \\
\XLSARHostedRow GPT-6.1-sol & -- & ZH & 15.30 & 6.24 & 2.39 & 0.88 & \underline{23.66} & 11.86 \\
Qwen2.5-VL & Qwen2.5-7B & ZH & \underline{17.75} & \underline{6.75} & 2.50 & 1.00 & 18.31 & \textbf{13.71} \\
LLaVA-OV & Qwen2-7B & ZH & \textbf{22.14} & \textbf{8.38} & \textbf{3.34} & \textbf{1.37} & 14.35 & \underline{13.00} \\
LLaVA-1.5 & Vicuna-v1.5-7B & ZH & 15.70 & 4.84 & 1.65 & 0.58 & 11.68 & 10.96 \\
InternVL3 & Qwen2.5-7B & ZH & 16.60 & 6.53 & \underline{2.59} & \underline{1.07} & 17.02 & 12.19 \\
GeoChat & Vicuna-v1.5-7B & ZH & 12.00 & 3.80 & 1.30 & 0.46 & 10.10 & 9.23 \\
\bottomrule
\end{tabularx}
\end{table*}

\section{Experimental Evaluation}
\label{sec:experiments}

\subsection{Experimental Setup}
\label{subsec:experimental_setup}

The MLLMs evaluated on XLSAR-Bench are grouped into three categories: (a) open-weight general-purpose MLLMs, including Qwen2.5-VL~\cite{bai2025qwen25vl}, LLaVA-OV~\cite{li2024llavaonevision}, LLaVA-1.5~\cite{liu2024improved}, and InternVL3~\cite{zhu2025internvl3}; (b) proprietary MLLMs, including GLM-5.3-flash~\cite{glm53flash2026} and GPT-6.1-sol~\cite{gpt61sol2026}; and (c) specialized models, including the RS-oriented GeoChat~\cite{kuckreja2024geochat} and the SAR-specific SARChat~\cite{ma2025sarchat}, which is built on the GLM-Edge-4B-Chat backbone. For fair comparison, we adopt a zero-shot setting with semantically equivalent prompts in English and Chinese. All models are evaluated on the same held-out samples using their supported image interfaces. We additionally introduce \textit{XLSAR-Baseline}, which applies standard LoRA~\cite{hu2022lora} to Qwen3.5-9B~\cite{qwen352026} using downsampled whole-image views without tiling or cropping. Adaptation and evaluation use disjoint source-scene partitions. Model configurations, implementation details, and additional settings are provided in the Appendix.

\subsection{Evaluation Strategy}
\label{subsec:evaluation_metrics}

For VQA, we evaluate seven sub-tasks using accuracy. Following the evaluation hierarchy of XLRS-Bench~\cite{wang2025xlrs}, we report L2 capability scores and the overall VQA average, with L3 task results available for detailed analysis. Unlike an unweighted average over capabilities, our VQA average is computed over all question instances. For OLUC, which may contain multiple correct options, predictions are considered correct only when they exactly match the reference answer. For visual grounding, we evaluate FVG and CBVG using mean intersection over union (mIoU) and accuracy at IoU thresholds of 0.5 and 0.7. Invalid bounding boxes receive zero IoU. For DIC, we adopt BLEU-1 to BLEU-4, METEOR, and ROUGE-L. Results are reported separately for English and Chinese, with all scores expressed as percentages.

\subsection{Main Results}
\label{subsec:zero_shot_results}

\noindent\textbf{Results on VQA tasks.}
Table~\ref{tab:zeroshot_vqa} compares the perception and reasoning capabilities of different MLLMs on XLSAR-Bench. The highest VQA averages reach 67.45\% and \BestZeroShot{} in English and Chinese, respectively. Nevertheless, model performance varies considerably across tasks. From these results, we draw two main observations: (1) \textit{Uneven Perception and Reasoning Abilities:} Models performing well on scene classification do not necessarily achieve comparable results on counting or spatial relationships, indicating that global scene recognition does not guarantee fine-grained understanding. (2) \textit{Challenges in Large-scene SAR Interpretation:} Tasks requiring accurate identification, selection, and counting of local targets remain difficult, highlighting the limitations of current MLLMs in capturing detailed visual information across large SAR scenes. Figure~\ref{fig:qual_vqa} presents one case per sub-task. On the two land-use tasks, Qwen2.5-VL and InternVL3 select wrong option subsets and LLaVA-OV returns an unsupported all-options set; on RC, Qwen2.5-VL misses the targets inside the marked region; on OSR and CCR all three models fail, and only OC is answered correctly by all of them. The per-case error categories are indexed in Table~\ref{tab:error_case}.

\begin{figure*}[!tb]
  \centering
  \begin{subfigure}[c]{0.495\textwidth}
    \includegraphics[width=\linewidth]{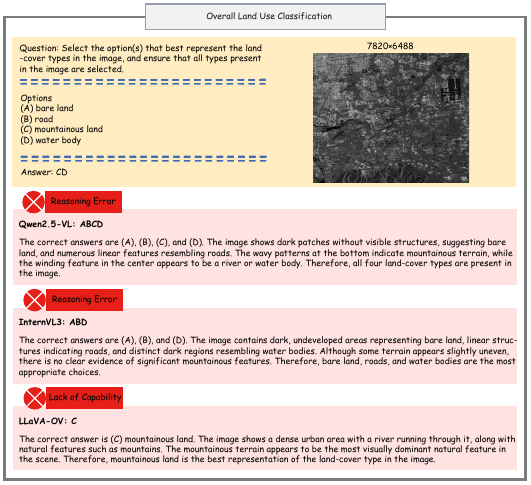}
    \caption{}\label{fig:vqa_a}
  \end{subfigure}\hspace{1pt}%
  \begin{subfigure}[c]{0.495\textwidth}
    \includegraphics[width=\linewidth]{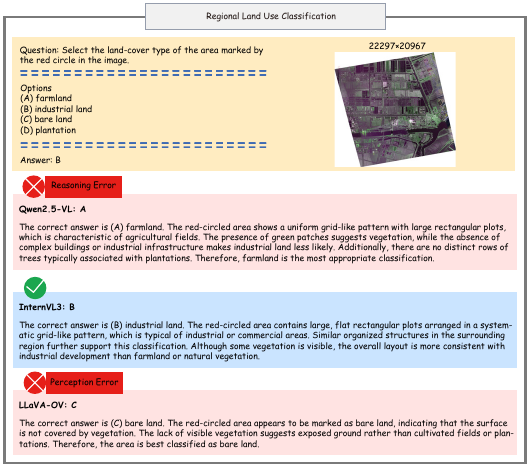}
    \caption{}\label{fig:vqa_b}
  \end{subfigure}

  \vspace{2pt}

  \begin{subfigure}[c]{0.495\textwidth}
    \includegraphics[width=\linewidth]{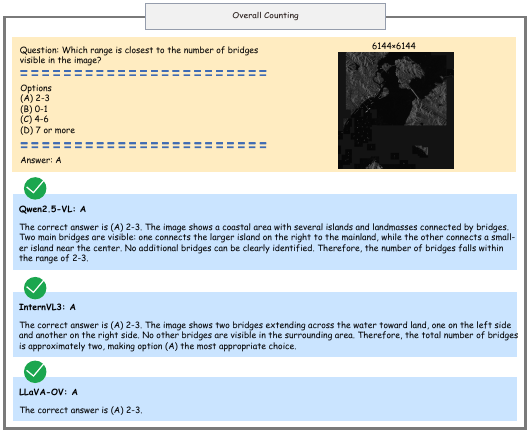}
    \caption{}\label{fig:vqa_c}
  \end{subfigure}\hspace{1pt}%
  \begin{subfigure}[c]{0.495\textwidth}
    \includegraphics[width=\linewidth]{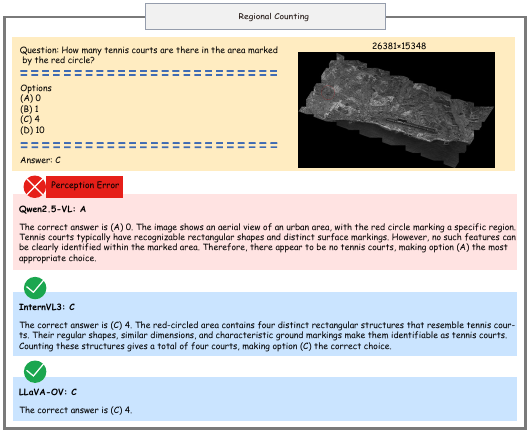}
    \caption{}\label{fig:vqa_d}
  \end{subfigure}
  \caption{Qualitative VQA cases of Qwen2.5-VL, InternVL3, and LLaVA-OV on the seven VQA sub-tasks (continued on the next page): (a)~OLUC, (b)~RLUC, (c)~OC, and (d)~RC. Each panel shows the complete scene, the question with options, the reference answer, and the three model outputs.}
  \label{fig:qual_vqa}
\end{figure*}

\begin{figure*}[!tb]
  \ContinuedFloat
  \centering
  \begin{subfigure}[c]{0.495\textwidth}
    \includegraphics[width=\linewidth]{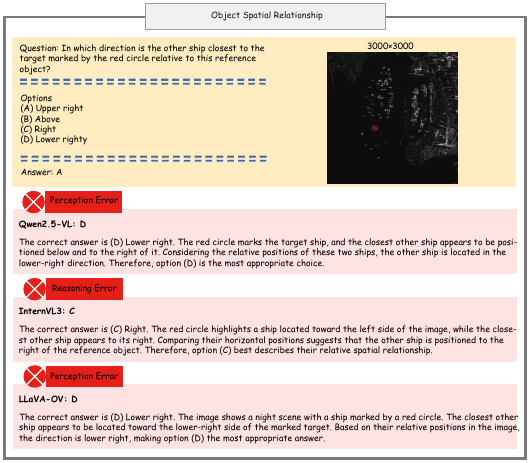}
    \caption{}\label{fig:vqa_e}
  \end{subfigure}\hspace{1pt}%
  \begin{subfigure}[c]{0.495\textwidth}
    \includegraphics[width=\linewidth]{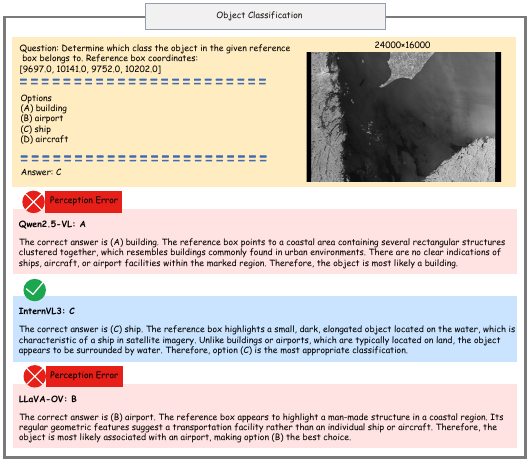}
    \caption{}\label{fig:vqa_f}
  \end{subfigure}

  \vspace{2pt}

  \begin{subfigure}[c]{0.495\textwidth}
    \includegraphics[width=\linewidth]{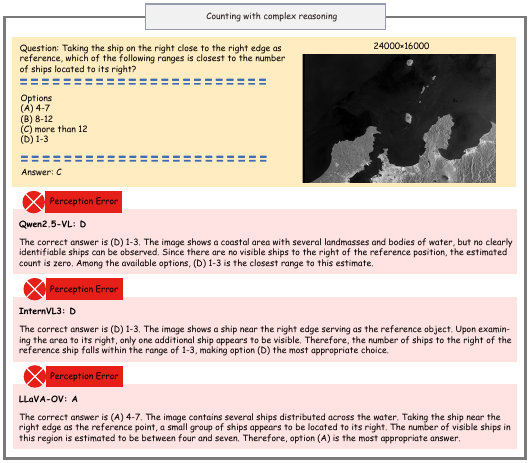}
    \caption{}\label{fig:vqa_g}
  \end{subfigure}\hspace{1pt}%
  \begin{subfigure}[c]{0.495\textwidth}
    \includegraphics[width=\linewidth]{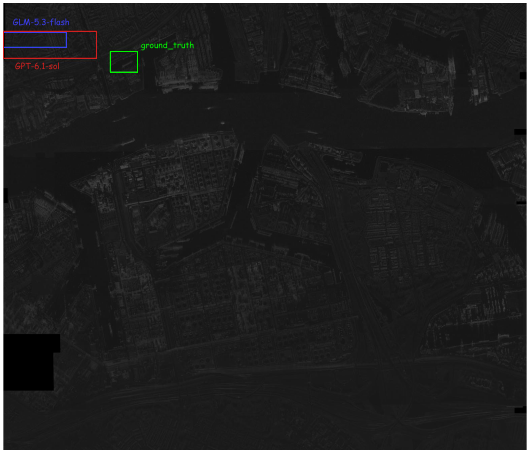}
    \caption{}\label{fig:vqa_h}
  \end{subfigure}
  \caption{Qualitative cases (continued): (e)~OSR, (f)~OCLS, (g)~CCR, and (h)~an FVG grounding case.}
\end{figure*}

\begin{table*}[!t]
\centering
\caption{Zero-shot visual-grounding results on XLSAR-Bench in EN/ZH. mIoU and accuracy at IoU 0.5/0.7 (A@.5, A@.7) are reported in percent. Proprietary models are highlighted in gray. The best score is in \textbf{bold} and the second-best is \underline{underlined}.}
\label{tab:zeroshot_grounding}
\XLSARResultTable
\begin{tabular}{@{}L{3.60cm}L{3.30cm}C{1.10cm}|C{1.49cm}C{1.49cm}C{1.49cm}|C{1.49cm}C{1.49cm}C{1.49cm}@{}}
\toprule
\multirow{2}{*}{Method} & \multirow{2}{*}{LLM} & \multirow{2}{*}{Lang.} &
\multicolumn{3}{c}{FVG} & \multicolumn{3}{c}{CBVG} \\
\cmidrule(lr){4-6}\cmidrule(lr){7-9}
& & & mIoU & A@.5 & A@.7 & mIoU & A@.5 & A@.7 \\
\midrule
\XLSARHostedRow GLM-5.3-flash & -- & EN & \underline{0.67} & \underline{0.21} & 0.00 & \underline{0.55} & \underline{0.08} & 0.00 \\
\XLSARHostedRow GPT-6.1-sol & -- & EN & \textbf{10.98} & \textbf{10.17} & \textbf{2.79} & \textbf{6.67} & \textbf{6.64} & \textbf{3.80} \\
Qwen2.5-VL & Qwen2.5-7B & EN & 0.07 & 0.00 & 0.00 & 0.05 & 0.00 & 0.00 \\
LLaVA-OV & Qwen2-7B & EN & 0.15 & 0.00 & 0.00 & 0.20 & 0.00 & 0.00 \\
LLaVA-1.5 & Vicuna-v1.5-7B & EN & 0.24 & 0.05 & 0.00 & 0.25 & 0.00 & 0.00 \\
InternVL3 & Qwen2.5-7B & EN & 0.14 & 0.00 & 0.00 & 0.15 & 0.00 & 0.00 \\
GeoChat & Vicuna-v1.5-7B & EN & 0.17 & 0.00 & 0.00 & 0.18 & 0.00 & 0.00 \\
SARChat & GLM-Edge-4B-Chat & EN & 0.12 & 0.00 & 0.00 & 0.22 & 0.00 & 0.00 \\
\midrule
\XLSARHostedRow GLM-5.3-flash & -- & ZH & \underline{1.16} & \underline{0.52} & \underline{0.05} & \underline{0.71} & \underline{0.38} & 0.00 \\
\XLSARHostedRow GPT-6.1-sol & -- & ZH & \textbf{11.62} & \textbf{10.38} & \textbf{3.75} & \textbf{8.36} & \textbf{8.35} & \textbf{3.80} \\
Qwen2.5-VL & Qwen2.5-7B & ZH & 0.10 & 0.00 & 0.00 & 0.06 & 0.00 & 0.00 \\
LLaVA-OV & Qwen2-7B & ZH & 0.14 & 0.00 & 0.00 & 0.20 & 0.00 & 0.00 \\
LLaVA-1.5 & Vicuna-v1.5-7B & ZH & 0.21 & 0.00 & 0.00 & 0.22 & 0.08 & 0.00 \\
InternVL3 & Qwen2.5-7B & ZH & 0.16 & 0.00 & 0.00 & 0.16 & 0.00 & 0.00 \\
GeoChat & Vicuna-v1.5-7B & ZH & 0.20 & 0.00 & 0.00 & 0.21 & 0.00 & 0.00 \\
SARChat & GLM-Edge-4B-Chat & ZH & 0.10 & 0.00 & 0.00 & 0.12 & 0.00 & 0.00 \\
\bottomrule
\end{tabular}
\end{table*}

\begin{figure*}[!tb]
  \centering
  \begin{subfigure}[c]{0.49\textwidth}
    \includegraphics[width=\linewidth]{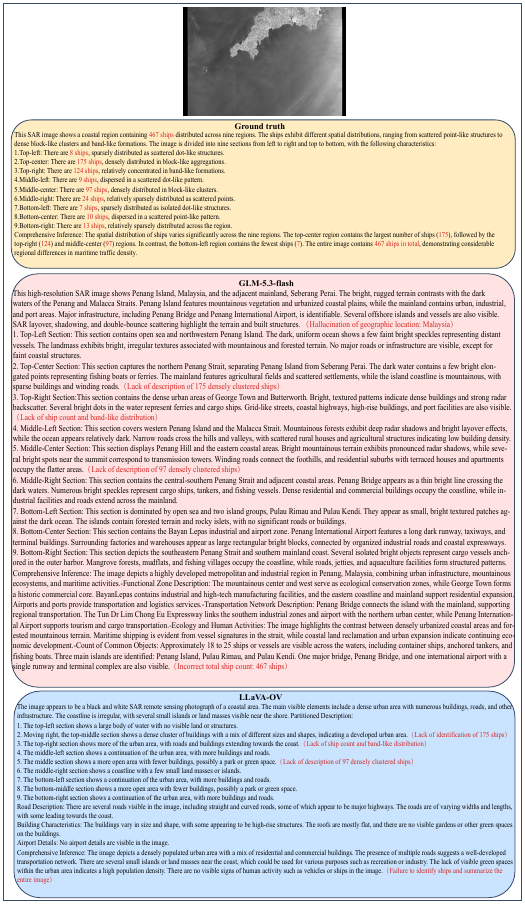}
    \caption{}\label{fig:dic_en}
  \end{subfigure}\hspace{4pt}%
  \begin{subfigure}[c]{0.49\textwidth}
    \includegraphics[width=\linewidth]{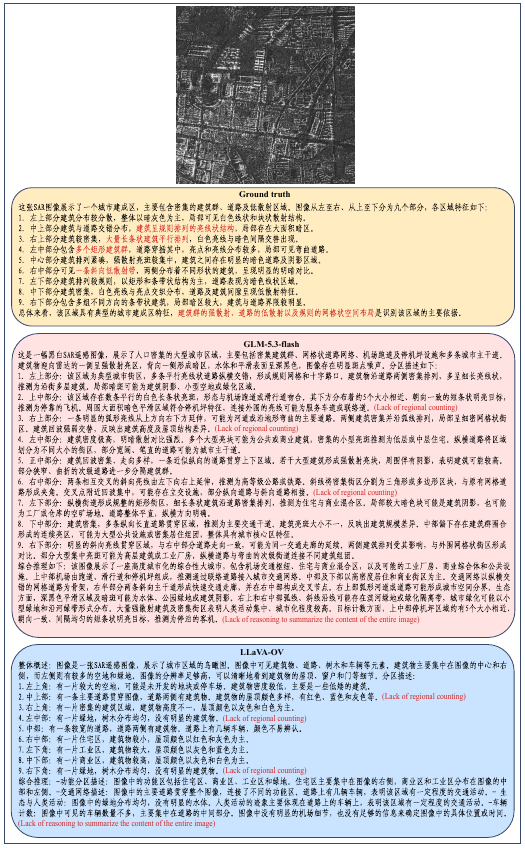}
    \caption{}\label{fig:dic_zh}
  \end{subfigure}
  \caption{DIC results of XLSAR-Bench in (a)~English and (b)~Chinese.}
  \label{fig:qual_dic}
\end{figure*}

\noindent\textbf{Results on image captioning tasks.}
Table~\ref{tab:zeroshot_caption} presents the DIC results. The highest METEOR scores reach 20.46\% and 24.04\% for English and Chinese, respectively, while BLEU-4 scores reach at most 2.13\% and 1.37\%. Qualitative analysis further reveals that MLLMs can often recognize the dominant scene but struggle to describe regional structures, small objects, and their spatial distributions. Some generated captions also contain unsupported details, indicating that comprehensive and accurate SAR scene description remains challenging. Figure~\ref{fig:qual_dic} shows two representative cases: in the English case the reference describes 467 ships across nine coastal regions, yet GLM-5.3-flash omits the regional ship distributions and LLaVA-OV describes only generic urban and coastal structures; in the Chinese case GLM-5.3-flash introduces unsupported airport facilities and LLaVA-OV misses the regional building distribution.

\noindent\textbf{Results on grounding tasks.}
Table~\ref{tab:zeroshot_grounding} reports the performance of MLLMs on FVG and CBVG. Both tasks present substantial challenges, particularly when targets occupy small areas or share similar backscatter characteristics with surrounding objects. CBVG further requires models to resolve semantic and spatial conditions before localization, resulting in generally weaker performance than FVG. These results reveal considerable limitations in fine-grained visual localization within large-scene SAR imagery. Figure~\ref{fig:qual_vqa}(h) shows a representative FVG case: asked to locate the horizontal port between two land areas in the upper left, GLM-5.3-flash localizes a nearby region farther to the left and GPT-6.1-sol predicts a larger box over a similar incorrect area.

\begin{table}[!tb]
\centering
\caption{LoRA adaptation results on XLSAR-Bench in EN/ZH. Scores are reported in percent: VQA Avg. is pooled accuracy, FVG/CBVG use mIoU, and DIC uses METEOR. The best score is in \textbf{bold} and the second-best is \underline{underlined}.}
\label{tab:adaptation_main}
\XLSARResultTable
\begin{tabular}{@{}L{3.30cm}C{0.60cm}|C{1.00cm}C{0.80cm}C{0.85cm}|C{0.90cm}@{}}
\toprule
Method & Lang. & VQA Avg. & \multicolumn{2}{c}{Grounding} & DIC \\
\cmidrule(lr){4-5}
& & & FVG & CBVG & \\
\midrule
Whole-image zero-shot & EN & \underline{33.67} & \textbf{0.14} & \textbf{0.20} & \underline{15.36} \\
\textbf{XLSAR-Baseline} (LoRA) & EN & \textbf{74.73} & \underline{0.13} & \underline{0.12} & \textbf{18.71} \\
\midrule
Whole-image zero-shot & ZH & \underline{35.89} & \textbf{0.16} & \textbf{0.02} & \underline{17.83} \\
\textbf{XLSAR-Baseline} (LoRA) & ZH & \textbf{75.54} & \underline{0.08} & \underline{0.00} & \textbf{19.28} \\
\bottomrule
\end{tabular}
\end{table}

\noindent\textbf{Results on whole-image adaptation.}
Table~\ref{tab:adaptation_main} compares XLSAR-Baseline with its untuned backbone under the same whole-image input setting. LoRA adaptation improves the VQA averages by 41.1 and 39.7 percentage points in English and Chinese, respectively. However, the improvements differ across tasks: VQA benefits substantially, captioning improves more moderately, and grounding performance remains limited. These findings suggest that straightforward domain adaptation can improve SAR semantic understanding but is insufficient to address fine-grained localization.

\section{Discussion}
\label{sec:discussion}

Despite improvements through adaptation, understanding large-scene SAR imagery remains challenging. First, downsampling complete scenes can obscure small targets, limiting fine-grained recognition and localization. Second, repeated objects with similar scattering characteristics introduce difficulties in counting and spatial reasoning. Our scale analysis on XLSAR-Baseline further shows a 5.94-percentage-point difference in VQA performance across image-size quartiles, although grounding results remain too low to establish a clear target-size trend. These observations highlight the need for MLLMs that better preserve local visual details while maintaining global scene understanding.

\section{Conclusion}
\label{sec:conclusion}

In this paper, we introduced XLSAR-Bench, a benchmark for systematically evaluating multimodal large language models on large-scene synthetic aperture radar imagery. XLSAR-Bench covers seven VQA sub-tasks together with visual grounding and DIC, supports both English and Chinese, and incorporates human-verified annotations. Through the evaluation of general-purpose and SAR-oriented MLLMs, we identified several important limitations in current large-scene SAR understanding. First, strong overall VQA performance can conceal substantial weaknesses in global scene interpretation and spatially constrained reasoning. Second, fine-grained visual grounding remains particularly challenging, with a pronounced performance gap between proprietary and open-weight models. Third, LoRA adaptation substantially improves VQA accuracy but does not yield corresponding improvements in grounding, highlighting the task-dependent nature of multimodal adaptation. Human evaluation and qualitative error analysis further demonstrate the difficulty of reliable counting, localization, and detailed scene description. These findings establish XLSAR-Bench as a multidimensional testbed for identifying the strengths and limitations of MLLMs in realistic large-scene SAR interpretation and for guiding future advances in spatially grounded multimodal understanding.

\appendices

\definecolor{macayellow}{RGB}{255,238,186}
\definecolor{macapink}{RGB}{255,214,224}
\definecolor{macablue}{RGB}{201,229,255}
\definecolor{macagreen}{RGB}{214,244,214}
\newcommand{\roundedboxyellow}[1]{%
  \tikz[baseline=(char.base)]{\node[anchor=south west, rounded corners, text height=1.5ex, text depth=.25ex, fill=macayellow, draw=none, text=black, font=\bfseries] (char) {#1};}}
\newcommand{\roundedboxred}[1]{%
  \tikz[baseline=(char.base)]{\node[anchor=south west, rounded corners, text height=1.5ex, text depth=.25ex, fill=macapink, draw=none, text=black, font=\bfseries] (char) {#1};}}
\newcommand{\roundedboxcyan}[1]{%
  \tikz[baseline=(char.base)]{\node[anchor=south west, rounded corners, text height=1.5ex, text depth=.25ex, fill=macablue, draw=none, text=black, font=\bfseries] (char) {#1};}}
\newcommand{\roundedboxgreen}[1]{%
  \tikz[baseline=(char.base)]{\node[anchor=south west, rounded corners, text height=1.5ex, text depth=.25ex, fill=macagreen, draw=none, text=black, font=\bfseries] (char) {#1};}}

\section{Human Evaluation}
\label{app-sec1}

\begin{figure}[!tb]
\centering
\includegraphics[width=\columnwidth]{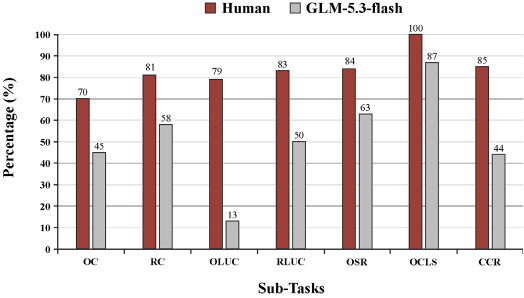}
\caption{Human evaluation results on XLSAR-Bench. Per-sub-task accuracy of two expert groups (averaged, 20 experts in total) and the hosted model GLM-5.3-flash.}
\label{fig:human}
\end{figure}

For all questions from each VQA sub-task (L3 dimensions), we had 20 experts, organized into two groups, to answer them simultaneously; the reported accuracy averages both groups. As shown in Figure~\ref{fig:human}, human accuracy stays between 70\% and 100\% across the seven sub-tasks, validating the reliability of XLSAR-Bench. Human evaluation is not error-free, however: the lowest human score appears on OC (70\%), confirming that counting over large scenes is inherently challenging even for humans. In contrast, the strongest hosted model in the main evaluation, GLM-5.3-flash, reaches its highest score on OCLS (87\%) and its lowest on OLUC (13\%); the resulting human--model gap exceeds 60 percentage points on OLUC.

\section{L3 Task Results}
\label{app-sec2}

Table~\ref{tab:zeroshot_vqa_l3} reports the per-task (L3) VQA results in both languages, which underlie the pooled L2 scores in the main text.

\begin{table*}[!tb]
\centering
\caption{Zero-shot L3 VQA results on XLSAR-Bench in EN/ZH. All results are reported in percent. Avg. denotes the pooled instance-level accuracy over all questions of the seven VQA sub-tasks rather than the unweighted average of sub-task accuracies. Proprietary models are highlighted in gray.}
\label{tab:zeroshot_vqa_l3}
\begin{threeparttable}
\footnotesize
\setlength{\tabcolsep}{2.2pt}
\renewcommand{\arraystretch}{1.08}
\begin{tabularx}{\textwidth}{@{}L{0.155\textwidth}L{0.135\textwidth}C{0.070\textwidth}YYYYYYYY@{}}
\toprule
\multirow{2}{*}{Method} &
\multirow{2}{*}{LLM} &
\multirow{2}{*}{Lang.} &
\multicolumn{6}{c}{Perception} &
\multicolumn{1}{c}{Reasoning} &
\multirow{2}{*}{Avg.} \\
\cmidrule(lr){4-9}\cmidrule(lr){10-10}
& & & OC & RC & OLUC & RLUC & OSR & OCLS & CCR & \\
\midrule
\rowcolor{gray!12} GLM-5.3-flash & -- & EN & 39.49 & \textbf{54.66} & 15.56 & 48.35 & \textbf{59.49} & \textbf{84.76} & \textbf{39.46} & \textbf{67.45} \\
\rowcolor{gray!12} GPT-6.1-sol & -- & EN & 36.28 & 50.92 & \textbf{25.93} & \textbf{52.66} & 47.70 & 80.82 & 38.26 & 63.26 \\
Qwen2.5-VL & Qwen2.5-7B & EN & 24.63 & 45.83 & 22.22 & 45.82 & 32.53 & 57.44 & 28.37 & 47.31 \\
LLaVA-OV & Qwen2-7B & EN & 22.89 & 44.12 & 8.15 & 38.73 & 30.41 & 58.61 & 29.48 & 46.90 \\
LLaVA-1.5 & Vicuna-v1.5-7B & EN & \textbf{45.52} & 24.79 & 5.19 & 25.57 & 28.78 & 22.56 & 24.12 & 25.03 \\
InternVL3 & Qwen2.5-7B & EN & 25.00 & 51.17 & 11.85 & 39.24 & 19.56 & 64.76 & 30.39 & 50.50 \\
GeoChat & Vicuna-v1.5-7B & EN & 43.95 & 24.53 & 17.04 & 25.57 & 34.02 & 21.91 & 34.01 & 26.05 \\
SARChat & GLM-Edge-4B-Chat & EN & 28.92 & 49.10 & 4.12 & 31.90 & 43.83 & 80.21 & 36.14 & 60.77 \\
\midrule
\rowcolor{gray!12} GLM-5.3-flash & -- & ZH & 45.21 & \textbf{57.95} & 12.69 & 49.36 & \textbf{63.25} & \textbf{86.56} & \textbf{43.97} & \textbf{70.17} \\
\rowcolor{gray!12} GPT-6.1-sol & -- & ZH & 44.44 & 48.56 & 17.04 & \textbf{50.89} & 48.07 & 82.70 & 39.83 & 64.10 \\
Qwen2.5-VL & Qwen2.5-7B & ZH & 30.25 & 45.62 & \textbf{20.74} & 48.50 & 36.78 & 70.79 & 30.87 & 54.83 \\
LLaVA-OV & Qwen2-7B & ZH & 29.59 & 48.50 & 12.59 & 38.23 & 29.63 & 71.68 & 29.11 & 54.56 \\
LLaVA-1.5 & Vicuna-v1.5-7B & ZH & \textbf{45.52} & 24.79 & 5.20 & 25.32 & 28.78 & 21.71 & 24.12 & 24.61 \\
InternVL3 & Qwen2.5-7B & ZH & 29.59 & 52.07 & 4.40 & 37.33 & 20.61 & 72.48 & 30.87 & 54.76 \\
GeoChat & Vicuna-v1.5-7B & ZH & \textbf{45.52} & 24.60 & 5.19 & 25.07 & 28.90 & 21.71 & 24.12 & 24.57 \\
SARChat & GLM-Edge-4B-Chat & ZH & 40.83 & 47.76 & 4.44 & 34.18 & 47.64 & 77.01 & 37.34 & 60.12 \\
\bottomrule
\end{tabularx}
\begin{tablenotes}[flushleft]
\scriptsize
\item[] OC/RC: Overall/Regional Counting; OLUC/RLUC: Overall/Regional Land Use Classification; OSR: Object Spatial Relationship; OCLS: Object Classification; CCR: Counting with Complex Reasoning. Bold marks the best score in each column within a language group.
\end{tablenotes}
\end{threeparttable}
\end{table*}

\begin{figure}[!t]
    \centering
    \includegraphics[width=\columnwidth]{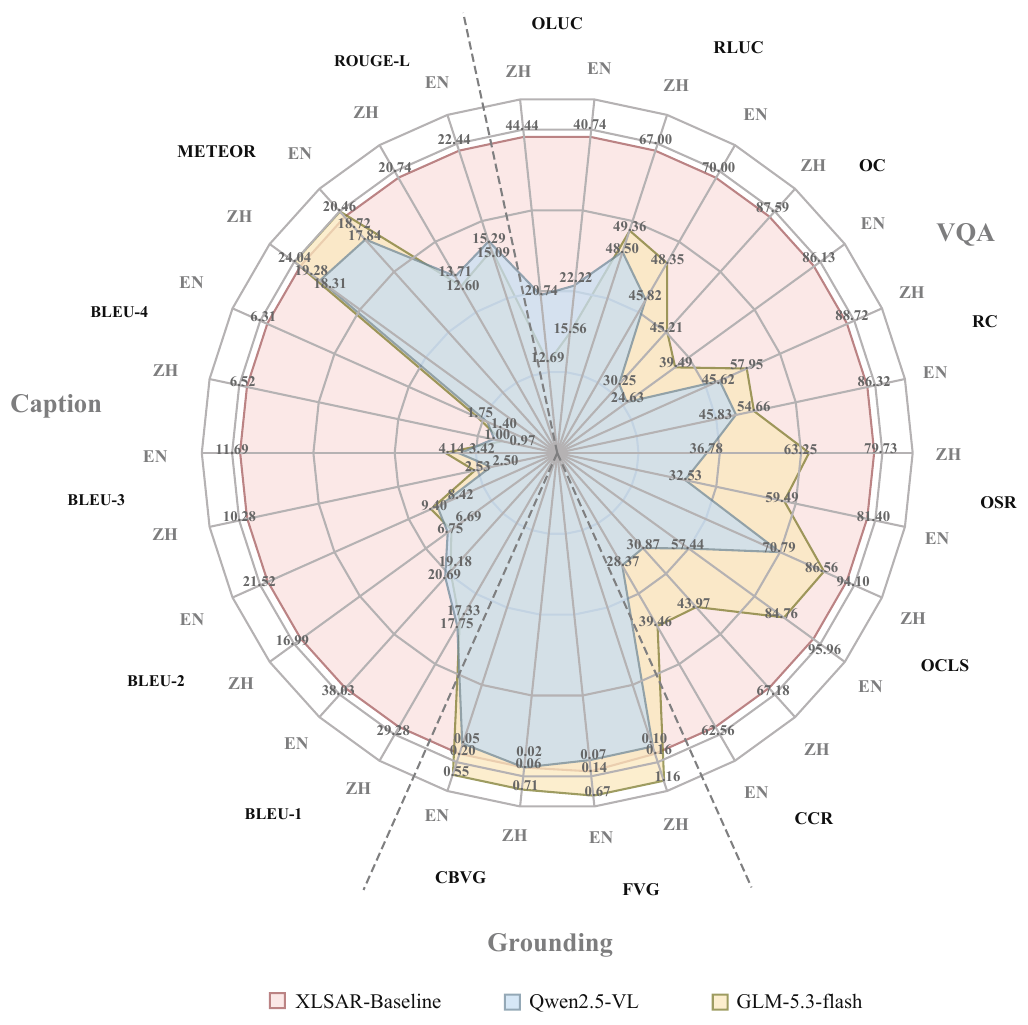}
    \caption{Task-level comparison between XLSAR-Baseline and the zero-shot Qwen2.5-VL and GLM-5.3-flash. Each axis reports the EN and ZH scores separately: OLUC, RLUC, OC, RC, OSR, OCLS, and CCR report accuracy; FVG and CBVG report mIoU; BLEU-1--4, METEOR, and ROUGE-L are reported for DIC.}
    \label{fig:radar}
\end{figure}

\noindent\textbf{Radar-chart overview.} Figure~\ref{fig:radar} contrasts the per-task capability profiles of the three models. The VQA axes expand markedly for XLSAR-Baseline after LoRA adaptation, whereas its grounding axes stay close to the center, mirroring the pooled results in Table~\ref{tab:adaptation_main}. GLM-5.3-flash leads most VQA axes (e.g., 84.76\%/86.56\% on EN/ZH OCLS and 48.35\%/49.36\% on EN/ZH RLUC) together with the captioning axes, while all three models collapse on the FVG and CBVG axes, where the best mIoU among them never exceeds 1.16. On every axis, the EN and ZH points trace nearly identical shapes, consistent with the small cross-language gaps in Table~\ref{tab:zeroshot_vqa_l3}.

\noindent\textbf{Marked-region cues separate localization from counting.} Every model scores higher on RLUC than on OLUC (by 8.5 to 32.8 points in EN), because the marked region removes the need to aggregate land-use evidence over the full scene. Weak open-weight models show the opposite pattern on counting: LLaVA-1.5 and GeoChat fall to 24.5\%--24.8\% on RC, essentially the 25\% chance level of the four-option format, while remaining above 43\% on OC, whose coarse count ranges tolerate imprecise global estimates. Their RC failure therefore stems from locating the marked region rather than from counting itself, identifying region-level grounding as the bottleneck capability. The cross-language gaps in Table~\ref{tab:zeroshot_vqa_l3} also track the LLM backbone: models built on a Qwen LLM gain 4.3 to 7.7 points in the ZH block, whereas the Vicuna-based models lose about one point.

\noindent\textbf{Grounding separates hosted from open-weight models most clearly.} On FVG, GPT-6.1-sol reaches an mIoU of 10.98--11.62 while the best open-weight model reaches only 0.24. Measured as ratios over the best open-weight score, hosted models lead by roughly 45$\times$ on grounding mIoU but only 1.1$\times$ on the pooled VQA average (67.45 vs.\ 60.77). Localization of small targets within complete large scenes is thus the capability that currently separates hosted models from open-weight ones most clearly.

\section{Error Taxonomy and Qualitative Cases}
\label{app-sec3}

Following the taxonomy of MMT-Bench~\cite{mmtbench2024}, three error categories are observed on XLSAR-Bench:

\noindent\roundedboxyellow{Perception Error}: the model fails to recognize, classify, or locate objects, mainly due to the limited representational power of its visual encoder; this is the most prevalent error type, especially for small or repeated targets.

\noindent\roundedboxred{Reasoning Error}: the model perceives the relevant visual content but fails at the reasoning step and returns an incorrect answer.

\noindent\roundedboxcyan{Lack of Capability}: the model does not possess the capability required by the task at all.

The qualitative cases of all seven VQA sub-tasks, together with the captioning and grounding qualitative results, are presented in the main text (Figs.~\ref{fig:qual_vqa} and~\ref{fig:qual_dic}). Table~\ref{tab:error_case} indexes the seven cases by sub-task and lists the error category of each model.

\begin{table*}[!tb]
\centering
\caption{Case-study panels of Fig.~\ref{fig:qual_vqa} by sub-task with the associated error categories for each MLLM.}
\label{tab:error_case}
\XLSARTableFont
\setlength{\tabcolsep}{3pt}
\resizebox{0.86\textwidth}{!}{%
\begin{tabular}{lllccc}
\toprule
Case panel & L2 task & L3 task & Qwen2.5-VL & InternVL3 & LLaVA-OV \\
\midrule
Fig.~\ref{fig:qual_vqa}(a) & Scene classification & Overall land use classification & \roundedboxred{Reasoning Error} & \roundedboxred{Reasoning Error} & \roundedboxcyan{Lack of Capability} \\
Fig.~\ref{fig:qual_vqa}(b) & Scene classification & Regional land use classification & \roundedboxred{Reasoning Error} & \roundedboxgreen{Correct} & \roundedboxyellow{Perception Error} \\
Fig.~\ref{fig:qual_vqa}(c) & Counting & Overall counting & \roundedboxgreen{Correct} & \roundedboxgreen{Correct} & \roundedboxgreen{Correct} \\
Fig.~\ref{fig:qual_vqa}(d) & Counting & Regional counting & \roundedboxyellow{Perception Error} & \roundedboxgreen{Correct} & \roundedboxgreen{Correct} \\
Fig.~\ref{fig:qual_vqa}(e) & Object spatial relationship & Object spatial relationship & \roundedboxyellow{Perception Error} & \roundedboxred{Reasoning Error} & \roundedboxyellow{Perception Error} \\
Fig.~\ref{fig:qual_vqa}(f) & Object properties & Object classification & \roundedboxyellow{Perception Error} & \roundedboxgreen{Correct} & \roundedboxyellow{Perception Error} \\
Fig.~\ref{fig:qual_vqa}(g) & Complex reasoning & Counting with complex reasoning & \roundedboxyellow{Perception Error} & \roundedboxyellow{Perception Error} & \roundedboxyellow{Perception Error} \\
\bottomrule
\end{tabular}}
\end{table*}

\bibliographystyle{IEEEtran}
\bibliography{reference} 

\begin{thebibliography}{10}
\providecommand{\url}[1]{#1}
\csname url@samestyle\endcsname
\providecommand{\newblock}{\relax}
\providecommand{\bibinfo}[2]{#2}
\providecommand{\BIBentrySTDinterwordspacing}{\spaceskip=0pt\relax}
\providecommand{\BIBentryALTinterwordstretchfactor}{4}
\providecommand{\BIBentryALTinterwordspacing}{\spaceskip=\fontdimen2\font plus
\BIBentryALTinterwordstretchfactor\fontdimen3\font minus
  \fontdimen4\font\relax}
\providecommand{\BIBforeignlanguage}[2]{{%
\expandafter\ifx\csname l@#1\endcsname\relax
\typeout{** WARNING: IEEEtran.bst: No hyphenation pattern has been}%
\typeout{** loaded for the language `#1'. Using the pattern for}%
\typeout{** the default language instead.}%
\else
\language=\csname l@#1\endcsname
\fi
#2}}
\providecommand{\BIBdecl}{\relax}
\BIBdecl

\bibitem{zhang2024earthgpt}
W.~Zhang, M.~Cai, T.~Zhang, Y.~Zhuang, and X.~Mao, ``{EarthGPT}: A universal
  multi-modal large language model for multi-sensor image comprehension in
  remote sensing domain,'' \emph{IEEE Trans. Geosci. Remote Sens.}, vol.~62, p.
  5917820, 2024.

\bibitem{zhang2025earthgptx}
W.~Zhang, M.~Cai, Y.~Ning, T.~Zhang, Y.~Zhuang, S.~Lu, H.~Chen, J.~Li, and
  X.~Mao, ``{EarthGPT-X}: A spatial {MLLM} for multilevel multisource remote
  sensing imagery understanding with visual prompting,'' \emph{IEEE Trans.
  Geosci. Remote Sens.}, vol.~63, pp. 1--21, 2025, art. no. 4709221.

\bibitem{wang2025xlrs}
F.~Wang, H.~Wang, Z.~Guo, D.~Wang, Y.~Wang, M.~Chen, Q.~Ma, L.~Lan, W.~Yang,
  J.~Zhang, Z.~Liu, and M.~Sun, ``{XLRS-Bench}: Could your multimodal {LLM}s
  understand extremely large ultra-high-resolution remote sensing imagery?'' in
  \emph{Proc. IEEE/CVF Conf. Comput. Vis. Pattern Recognit. (CVPR)}, Jun. 2025,
  pp. 14\,325--14\,336.

\bibitem{moreira2013sar_tutorial}
A.~Moreira, P.~Prats-Iraola, M.~Younis, G.~Krieger, I.~Hajnsek, and K.~P.
  Papathanassiou, ``A tutorial on synthetic aperture radar,'' \emph{IEEE
  Geosci. Remote Sens. Mag.}, vol.~1, no.~1, pp. 6--43, 2013.

\bibitem{argenti2013speckle}
F.~Argenti, A.~Lapini, T.~Bianchi, and L.~Alparone, ``A tutorial on speckle
  reduction in synthetic aperture radar images,'' \emph{IEEE Geosci. Remote
  Sens. Mag.}, vol.~1, no.~3, pp. 6--35, 2013.

\bibitem{zhu2021deep_sar}
X.~X. Zhu, S.~Montazeri, M.~Ali, Y.~Hua, Y.~Wang, L.~Mou, Y.~Shi, F.~Xu, and
  R.~Bamler, ``Deep learning meets {SAR}: Concepts, models, pitfalls, and
  perspectives,'' \emph{IEEE Geosci. Remote Sens. Mag.}, vol.~9, no.~4, pp.
  143--172, 2021.

\bibitem{ma2025sarchat}
Z.~Ma, X.~Xiao, S.~Dong, P.~Wang, H.~Wang, and Q.~Pan, ``{SARChat-Bench-2M}: A
  multi-task vision-language benchmark for {SAR} image interpretation,''
  \emph{arXiv preprint arXiv:2502.08168}, 2025.

\bibitem{wei2026sarlang}
Y.~Wei, A.~Xiao, Y.~Ren, Y.~Zhu, H.~Chen, J.~Xia, and N.~Yokoya,
  ``{SARLANG-1M}: A benchmark for vision-language modeling in {SAR} image
  understanding,'' \emph{IEEE Trans. Geosci. Remote Sens.}, vol.~64, pp. 1--20,
  2026, art. no. 5201320.

\bibitem{wang2026sareval}
Z.~Wang, L.~Liu, G.~Wan, Y.~Lu, F.~Zheng, G.~Sun, Y.~Huang, S.~Guo, X.~Li, and
  L.~Yuan, ``{SAREval}: A multi-dimensional and multi-task benchmark for
  evaluating visual language models on {SAR} image understanding,''
  \emph{Remote Sens.}, vol.~18, no.~1, p.~82, 2026.

\bibitem{li2025tacmt}
T.~Li, C.~Wang, S.~Tian, B.~Zhang, F.~Wu, Y.~Tang, and H.~Zhang, ``{TACMT}:
  Text-aware cross-modal transformer for visual grounding on high-resolution
  {SAR} images,'' \emph{ISPRS J. Photogramm. Remote Sens.}, vol. 222, pp.
  152--166, 2025.

\bibitem{ma2026sarvlm}
Q.~Ma, X.~Lu, W.~Liu, P.~Duan, X.~Kang, and S.~Li, ``{SARVLM}: A vision
  language foundation model for semantic understanding in {SAR} imagery,''
  \emph{arXiv preprint arXiv:2510.22665}, 2025.

\bibitem{gaosarcap2025}
Z.~Gao, S.~Sun, M.-M. Cheng, Y.~Liu, and L.~Liu, ``Multimodal large models
  driven {SAR} image captioning: A benchmark dataset and baselines,''
  \emph{IEEE J. Sel. Topics Appl. Earth Observ. Remote Sens.}, vol.~18, pp.
  24\,011--24\,026, 2025.

\bibitem{li2023pope}
Y.~Li, Y.~Du, K.~Zhou, J.~Wang, X.~Zhao, and J.-R. Wen, ``Evaluating object
  hallucination in large vision-language models,'' in \emph{Proc. Conf.
  Empirical Methods Natural Language Processing (EMNLP)}, 2023, pp. 292--305.

\bibitem{parmar2023instructionbias}
M.~Parmar, S.~Mishra, M.~Geva, and C.~Baral, ``Don't blame the annotator: Bias
  already starts in the annotation instructions,'' in \emph{Proc. 17th Conf.
  Eur. Chapter Assoc. Comput. Linguistics (EACL)}, 2023, pp. 1779--1789.

\bibitem{ls_ssdd}
T.~Zhang, X.~Zhang, X.~Ke, X.~Zhan, J.~Shi, S.~Wei, D.~Pan, J.~Li, H.~Su,
  Y.~Zhou, and D.~Kumar, ``{LS-SSDD-v1.0}: A deep learning dataset dedicated to
  small ship detection from large-scale {Sentinel-1} {SAR} images,''
  \emph{Remote Sens.}, vol.~12, no.~18, p. 2997, 2020.

\bibitem{sad}
D.~Wang, F.~Zhang, F.~Ma, W.~Hu, Y.~Tang, and Y.~Zhou, ``A benchmark
  {Sentinel-1} {SAR} dataset for airport detection,'' \emph{IEEE J. Sel. Topics
  Appl. Earth Observ. Remote Sens.}, vol.~15, pp. 6671--6686, 2022.

\bibitem{air_sarship}
X.~Sun, Z.~Wang, Y.~Sun, W.~Diao, Y.~Zhang, and K.~Fu, ``{AIR-SARShip-1.0}:
  High-resolution {SAR} ship detection dataset,'' \emph{J. Radars}, vol.~8,
  no.~6, pp. 852--862, 2019.

\bibitem{msar_bridge}
\BIBentryALTinterwordspacing
J.~Chen, Z.~Huang, R.~Xia, B.~Wu, L.~Sheng, L.~Sun, and B.~Yao, ``Large-scale
  multi-class {SAR} image target detection dataset-1.0 ({MSAR-1.0}),'' Journal
  of Radars online dataset, 2022, accessed July 22, 2026. [Online]. Available:
  \url{https://radars.ac.cn/web/data/getData?dataType=MSAR_en&pageType=en}
\BIBentrySTDinterwordspacing

\bibitem{whu_opt_sar}
X.~Li, G.~Zhang, H.~Cui, S.~Hou, S.~Wang, X.~Li, Y.~Chen, Z.~Li, and L.~Zhang,
  ``{MCANet}: A joint semantic segmentation framework of optical and {SAR}
  images for land use classification,'' \emph{Int. J. Appl. Earth Observ.
  Geoinf.}, vol. 106, p. 102638, 2022.

\bibitem{air_polsar_seg}
Z.~Wang, L.~Zhao, Y.~Wang, X.~Zeng, J.~Kang, J.~Yang, and X.~Sun,
  ``{AIR-PolSAR-Seg-2.0}: Polarimetric {SAR} ground terrain classification
  dataset for large-scale complex scenes,'' \emph{J. Radars}, vol.~14, no.~2,
  pp. 353--365, 2025.

\bibitem{spacenet6}
J.~Shermeyer, D.~Hogan, J.~Brown, A.~Van~Etten, N.~Weir, F.~Pacifici,
  R.~H{\"a}nsch, A.~Bastidas, S.~Soenen, T.~M. Bacastow, and R.~Lewis,
  ``{SpaceNet 6}: Multi-sensor all weather mapping dataset,'' in \emph{Proc.
  IEEE/CVF Conf. Comput. Vis. Pattern Recognit. Workshops (CVPRW)}, 2020, pp.
  768--777.

\bibitem{fusar_map}
X.~Shi, S.~Fu, J.~Chen, F.~Wang, and F.~Xu, ``Object-level semantic
  segmentation on the high-resolution {Gaofen-3 FUSAR-Map} dataset,''
  \emph{IEEE J. Sel. Topics Appl. Earth Observ. Remote Sens.}, vol.~14, pp.
  3107--3119, 2021.

\bibitem{yeseg}
K.~Wei, J.~Dai, D.~Hong, and Y.~Ye, ``{MGFNet}: An {MLP}-dominated gated fusion
  network for semantic segmentation of high-resolution multi-modal remote
  sensing images,'' \emph{Int. J. Appl. Earth Observ. Geoinf.}, vol. 135, p.
  104241, 2024.

\bibitem{atrnetstar2026}
Y.~Liu, W.~Li, L.~Liu, J.~Zhou, B.~Peng, Y.~Song, X.~Xiong, W.~Yang, T.~Liu,
  Z.~Liu, and X.~Li, ``{ATRNet-STAR}: A large dataset and benchmark towards
  remote sensing object recognition in the wild,'' \emph{IEEE Trans. Pattern
  Anal. Mach. Intell.}, pp. 1--18, 2026.

\bibitem{zhang2024mmerealworld}
Y.-F. Zhang, H.~Tian, W.~Wang \emph{et~al.}, ``{MME-RealWorld}: Could your
  multimodal {LLM} challenge high-resolution real-world scenarios that are hard
  to comprehend,'' in \emph{Proc. Int. Conf. Learn. Represent. (ICLR)}, 2025.

\bibitem{lu2018rsicd}
X.~Lu, B.~Wang, X.~Zheng, and X.~Li, ``Exploring models and data for remote
  sensing image caption generation,'' \emph{IEEE Trans. Geosci. Remote Sens.},
  vol.~56, no.~4, pp. 2183--2195, 2018.

\bibitem{lobry2020rsvqa}
S.~Lobry, D.~Marcos, J.~Murray, and D.~Tuia, ``{RSVQA}: Visual question
  answering for remote sensing data,'' \emph{IEEE Trans. Geosci. Remote Sens.},
  vol.~58, no.~12, pp. 8555--8566, 2020.

\bibitem{zhan2023rsvg}
Y.~Zhan, Z.~Xiong, and Y.~Yuan, ``{RSVG}: Exploring data and models for visual
  grounding on remote sensing data,'' \emph{IEEE Trans. Geosci. Remote Sens.},
  vol.~61, pp. 1--13, 2023.

\bibitem{li2024vrsbench}
X.~Li, J.~Ding, and M.~Elhoseiny, ``{VRSBench}: A versatile vision-language
  benchmark dataset for remote sensing image understanding,'' in \emph{Proc.
  Adv. Neural Inf. Process. Syst. (NeurIPS)}, vol.~37, 2024, pp. 3229--3242.

\bibitem{saratrx2025}
W.~Li, W.~Yang, Y.~Hou, L.~Liu, Y.~Liu, and X.~Li, ``{SARATR-X}: Toward
  building a foundation model for {SAR} target recognition,'' \emph{IEEE Trans.
  Image Process.}, vol.~34, pp. 869--884, 2025.

\bibitem{weng2026clears}
X.~Weng, R.~Ni, C.~Pang, X.~Hao, Y.~Wang, X.~Zhang, W.~Xu, and G.-S. Xia,
  ``Continual vision-language learning for remote sensing: Benchmarking and
  analysis,'' \emph{arXiv preprint arXiv:2604.00820}, 2026.

\bibitem{he2025sartext}
Y.~He, X.~Cheng, J.~Zhu, C.~Qiu, J.~Wang, X.~Zhang, Q.~Huang, and K.~Yang,
  ``{SAR-TEXT}: A large-scale {SAR} image-text dataset built with
  {SAR-Narrator} and progressive transfer learning,'' \emph{arXiv preprint
  arXiv:2507.18743}, 2025.

\bibitem{zhang2026fsarcap}
J.~Zhang, L.~Cong, D.~Zhuang, L.~Zhang, and B.~Zou, ``{FSAR-Cap}: A
  fine-grained two-stage annotated dataset for {SAR} image captioning,''
  \emph{IEEE Geosci. Remote Sens. Lett.}, vol.~23, 2026.

\bibitem{jiang2025sarclip}
C.~Jiang, C.~Wang, F.~Wu, P.~Ma, L.~Zou, T.~Li, J.~Ning, and Y.~Tang,
  ``{SARCLIP}: A multimodal foundation framework for {SAR} imagery via
  contrastive language-image pre-training,'' \emph{ISPRS J. Photogramm. Remote
  Sens.}, 2026.

\bibitem{ssicd2022}
K.~Zhao and W.~Xiong, ``Exploring data and models in {SAR} ship image
  captioning,'' \emph{IEEE Access}, vol.~10, pp. 91\,150--91\,159, 2022.

\bibitem{li2025hrssrd}
Y.~Li, W.~Liu, W.~Lu, R.~Li, F.~Li, T.~Huan, S.~Wang, and Y.~Wu, ``Synthetic
  aperture radar image captioning: Building a dataset and explore models,'' in
  \emph{Proc. 5th Int. Conf. Neural Netw. Inf. Commun. Eng. (NNICE)}, 2025, pp.
  465--472.

\bibitem{zhan2025vgrss}
Y.~Chen, L.~Zhan, Y.~Zhao \emph{et~al.}, ``{VGRSS}: Datasets and models for
  visual grounding in remote sensing ship images,'' \emph{IEEE Trans. Geosci.
  Remote Sens.}, vol.~63, pp. 1--11, 2025, art. no. 4703411.

\bibitem{bui2026saavg}
D.~Bui, M.-K. Nguyen-Thi, and M.-T. Tran-Nguyen, ``Asymmetric hierarchical
  fusion for visual grounding in {SAR} images,'' in \emph{Proc. Asian Conf.
  Intell. Inf. Database Syst.}, ser. Lecture Notes in Computer Science, vol.
  16530.\hskip 1em plus 0.5em minus 0.4em\relax Springer, 2026, pp. 399--414.

\bibitem{tang2026optsar}
X.~Tang, J.~Dong, J.~Cheng, and R.~Fan, ``{OptiSAR-Net++}: A large-scale
  benchmark and transformer-free framework for cross-domain remote sensing
  visual grounding,'' \emph{arXiv preprint arXiv:2603.24876}, 2026.

\bibitem{wang2024qwen2vl}
P.~Wang, S.~Bai, S.~Tan, S.~Wang, Z.~Fan, J.~Bai \emph{et~al.}, ``{Qwen2-VL}:
  Enhancing vision-language model's perception of the world at any
  resolution,'' \emph{arXiv preprint arXiv:2409.12191}, 2024.

\bibitem{bai2025qwen25vl}
S.~Bai, K.~Chen, X.~Liu, J.~Wang, W.~Ge, S.~Song \emph{et~al.}, ``{Qwen2.5-VL}
  technical report,'' \emph{arXiv preprint arXiv:2502.13923}, 2025.

\bibitem{chen2024internvl15}
Z.~Chen, W.~Wang, H.~Tian, S.~Ye, Z.~Gao, E.~Cui \emph{et~al.}, ``How far are
  we to {GPT}-4v? closing the gap to commercial multimodal models with
  open-source suites,'' \emph{Sci. China Inf. Sci.}, vol.~67, no.~12, p.
  220101, 2024.

\bibitem{guo2024llavauhd}
Z.~Guo, R.~Xu, Y.~Yao, J.~Cui, Z.~Ni, C.~Ge, T.-S. Chua, Z.~Liu, and G.~Huang,
  ``{LLaVA-UHD}: An {LMM} perceiving any aspect ratio and high-resolution
  images,'' in \emph{Proc. Eur. Conf. Comput. Vis. (ECCV)}, 2024, pp. 390--406.

\bibitem{li2025tokenpacker}
W.~Li, Y.~Yuan, J.~Liu, D.~Tang, S.~Wang, J.~Qin, J.~Zhu, and L.~Zhang,
  ``Tokenpacker: Efficient visual projector for multimodal {LLM},'' \emph{Int.
  J. Comput. Vis.}, vol. 133, pp. 6794--6812, 2025.

\bibitem{luo2025largeRSI}
J.~Luo, Y.~Zhang, X.~Yang, K.~Wu, Q.~Zhu, L.~Liang, J.~Chen, and Y.~Li, ``When
  large vision-language model meets large remote sensing imagery:
  Coarse-to-fine text-guided token pruning,'' in \emph{Proc. IEEE/CVF Int.
  Conf. Comput. Vis. (ICCV)}, 2025, pp. 9206--9217.

\bibitem{liu2026zoomearth}
R.~Liu, B.~Fu, J.~Song, K.~Li, W.~Li, L.~Xue, H.~Qiao, W.~Zhang, D.~Meng, and
  X.~Cao, ``{ZoomEarth}: Active perception for ultra-high-resolution geospatial
  vision-language tasks,'' in \emph{Proc. IEEE/CVF Conf. Comput. Vis. Pattern
  Recognit. (CVPR)}, Jun. 2026, pp. 34\,877--34\,888.

\bibitem{kuckreja2024geochat}
K.~Kuckreja, M.~S. Danish, M.~Naseer, A.~Das, S.~Khan, and F.~S. Khan,
  ``{GeoChat}: Grounded large vision-language model for remote sensing,'' in
  \emph{Proc. IEEE/CVF Conf. Comput. Vis. Pattern Recognit. (CVPR)}, 2024, pp.
  27\,831--27\,840.

\bibitem{muhtar2024lhrsbot}
D.~Muhtar, Z.~Li, F.~Gu, X.~Zhang, and P.~Xiao, ``{LHRS-Bot}: Empowering remote
  sensing with vgi-enhanced large multimodal language model,'' in \emph{Proc.
  Eur. Conf. Comput. Vis. (ECCV)}, 2024, pp. 440--457.

\bibitem{hu2022lora}
E.~J. Hu, Y.~Shen, P.~Wallis, Z.~Allen-Zhu, Y.~Li, S.~Wang, L.~Wang, and
  W.~Chen, ``{LoRA}: Low-rank adaptation of large language models,'' in
  \emph{Proc. Int. Conf. Learn. Represent. (ICLR)}, 2022.

\bibitem{luo2023lavin}
G.~Luo, Y.~Zhou, T.~Ren, S.~Chen, X.~Sun, and R.~Ji, ``Cheap and quick:
  Efficient vision-language instruction tuning for large language models,'' in
  \emph{Proc. Adv. Neural Inf. Process. Syst. (NeurIPS)}, vol.~36, 2023.

\bibitem{hrsid}
S.~Wei, X.~Zeng, Q.~Qu, M.~Wang, H.~Su, and J.~Shi, ``{HRSID}: A
  high-resolution {SAR} images dataset for ship detection and instance
  segmentation,'' \emph{IEEE Access}, vol.~8, pp. 120\,234--120\,254, 2020.

\bibitem{earthmiss}
Y.~Zhou, A.~Ma, J.~Wang, Z.~Chen, and Y.~Zhong, ``Remote sensing meta modal
  representation for missing modality land cover mapping: From {EarthMiss}
  dataset to {MetaRS} method,'' \emph{Remote Sens. Environ.}, vol. 333, p.
  115132, 2026.

\bibitem{air_mdsar_map}
N.~Zheng, Z.~Yang, X.~Shi, H.~Yang, Y.~Sun, and F.~Wang, ``Airborne
  multidimensional {SAR} land cover dataset and fusion classification method,''
  \emph{Nat. Remote Sens. Bull.}, vol.~28, no.~9, pp. 2209--2222, 2024.

\bibitem{li2024llavaonevision}
B.~Li, Y.~Zhang, D.~Guo, R.~Zhang, F.~Li, H.~Zhang, K.~Zhang, P.~Li, Z.~Liu,
  and C.~Li, ``{LLaVA-OneVision}: Easy visual task transfer,'' \emph{arXiv
  preprint arXiv:2408.03326}, 2024.

\bibitem{liu2024improved}
H.~Liu, C.~Li, Y.~Li, and Y.~J. Lee, ``Improved baselines with visual
  instruction tuning,'' in \emph{Proc. IEEE/CVF Conf. Comput. Vis. Pattern
  Recognit. (CVPR)}, 2024, pp. 26\,296--26\,305.

\bibitem{zhu2025internvl3}
J.~Zhu, W.~Zhang, Z.~Chen, Z.~Liu, Y.~Ye, L.~Guo, Y.~Su, J.~Qin, S.~Zhang,
  J.~Ye \emph{et~al.}, ``{InternVL3}: Exploring advanced training and test-time
  reasoning in open-source multimodal models,'' \emph{arXiv preprint
  arXiv:2504.10479}, 2025.

\bibitem{glm53flash2026}
{Zhipu AI}, ``{GLM-5.3-flash} multimodal model via the {GLM} open platform,''
  Available: \url{https://bigmodel.cn} (accessed Nov. 2026), 2026.

\bibitem{gpt61sol2026}
{OpenAI}, ``{GPT-6.1-sol} api model documentation,'' Available:
  \url{https://platform.openai.com/docs/models} (accessed Nov. 2026), 2026.

\bibitem{qwen352026}
\BIBentryALTinterwordspacing
{Qwen Team}, ``{Qwen3.5-9B} model card,'' Hugging Face model release, 2026,
  accessed July 22, 2026. [Online]. Available:
  \url{https://huggingface.co/Qwen/Qwen3.5-9B}
\BIBentrySTDinterwordspacing

\bibitem{mmtbench2024}
K.~Ying, F.~Meng, J.~Wang, Z.~Li, H.~Lin, Y.~Yang, H.~Zhang, W.~Zhang, Y.~Lin,
  S.~Liu \emph{et~al.}, ``{MMT-Bench}: A comprehensive multimodal benchmark for
  evaluating large vision-language models towards multitask {AGI},'' in
  \emph{Proc. Int. Conf. Mach. Learn. (ICML)}, 2024, pp. 57\,116--57\,198.

\end{thebibliography}
\vspace{-3em}
\begin{IEEEbiography}[{\includegraphics[width=1in,height=1.25in,clip,keepaspectratio]{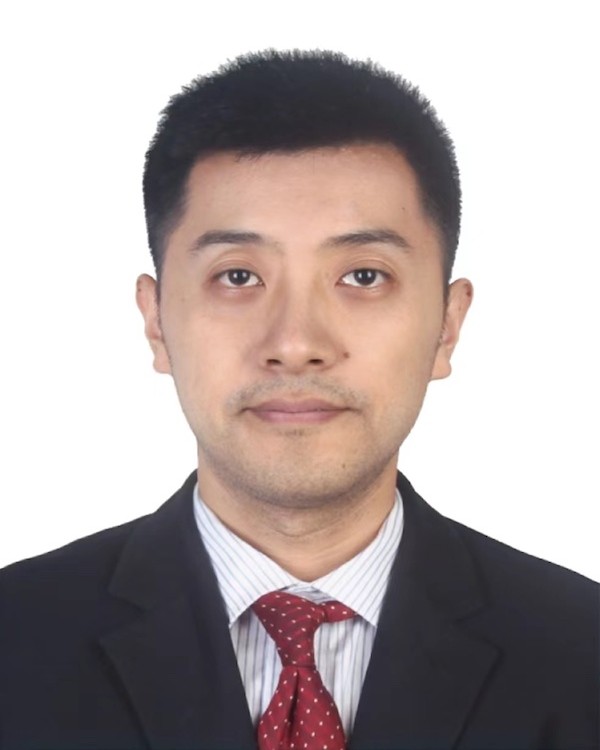}}]{La Jiang}
received the B.E. degree in electronic information engineering from Harbin Institute of Technology, Harbin, China, in 2006, and the M.S. degree in Integrated Circuit Design from Imperial College London, UK, in 2007.

He is currently working toward the Ph.D. degree in information and communication engineering with the College of Electronic Science and Technology, National University of Defense Technology (NUDT), Changsha, China. His research interests include remote-sensing image processing, multi-source data fusion and deep learning.
\end{IEEEbiography}
\vspace{-3em}
\begin{IEEEbiography}[{\includegraphics[width=1in,height=1.25in,clip,keepaspectratio]{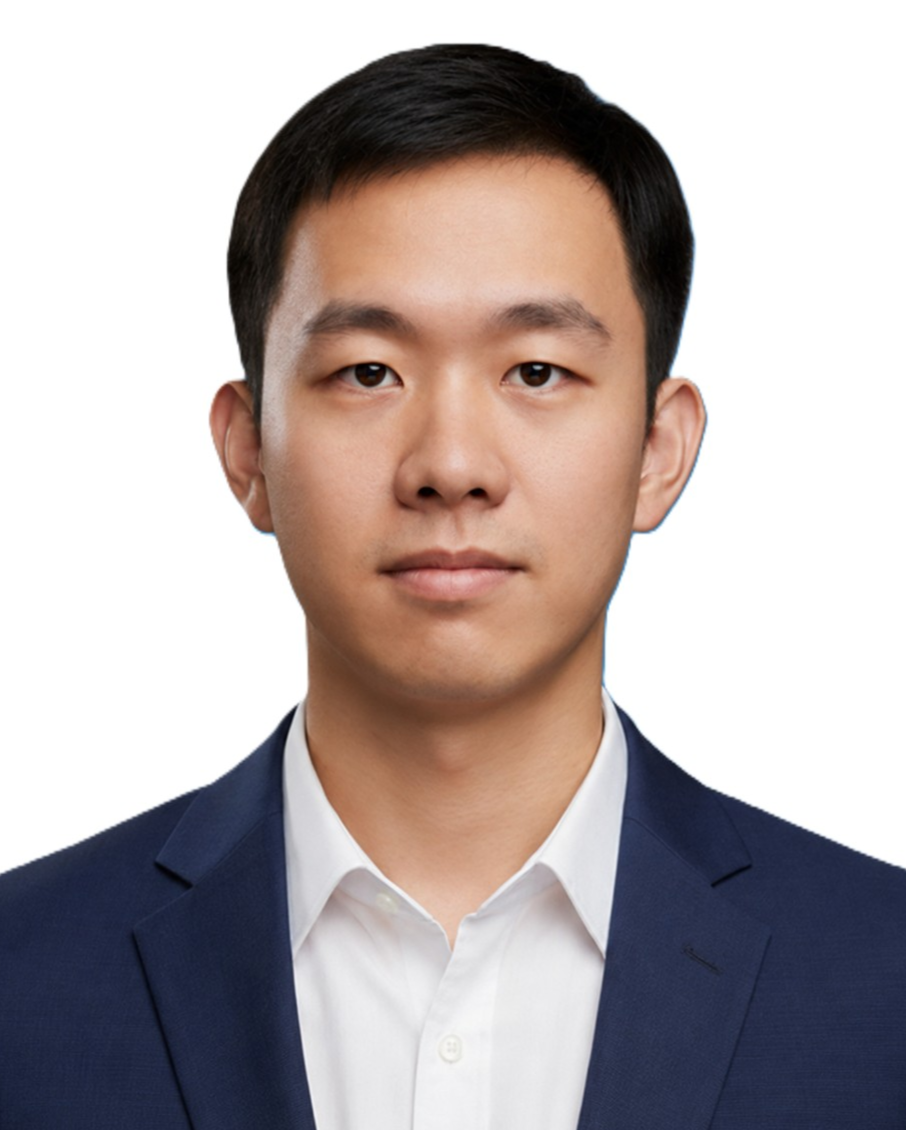}}]{Lingfeng Chen}
obtained the B.S. degree in microelectronic science and engineering at National University of Defense Technology (NUDT), Changsha, Hunan, China, in 2025.

He is currently pursuing a Ph.D. degree in information and communication engineering at NUDT. His research interest includes deep learning, meta-learning, foundation models and their application in radar automatic target recognition. He also serves as a reviewer for several journals, including IEEE Transactions on Geoscience and Remote Sensing, IEEE Transactions on Radar Systems, IEEE Signal Processing Letters, and IEEE Journal of Selected Topics in Applied Earth Observations.
\end{IEEEbiography}
\vspace{-3em}
\begin{IEEEbiography}[{\includegraphics[width=1in,height=1.25in,clip,keepaspectratio]{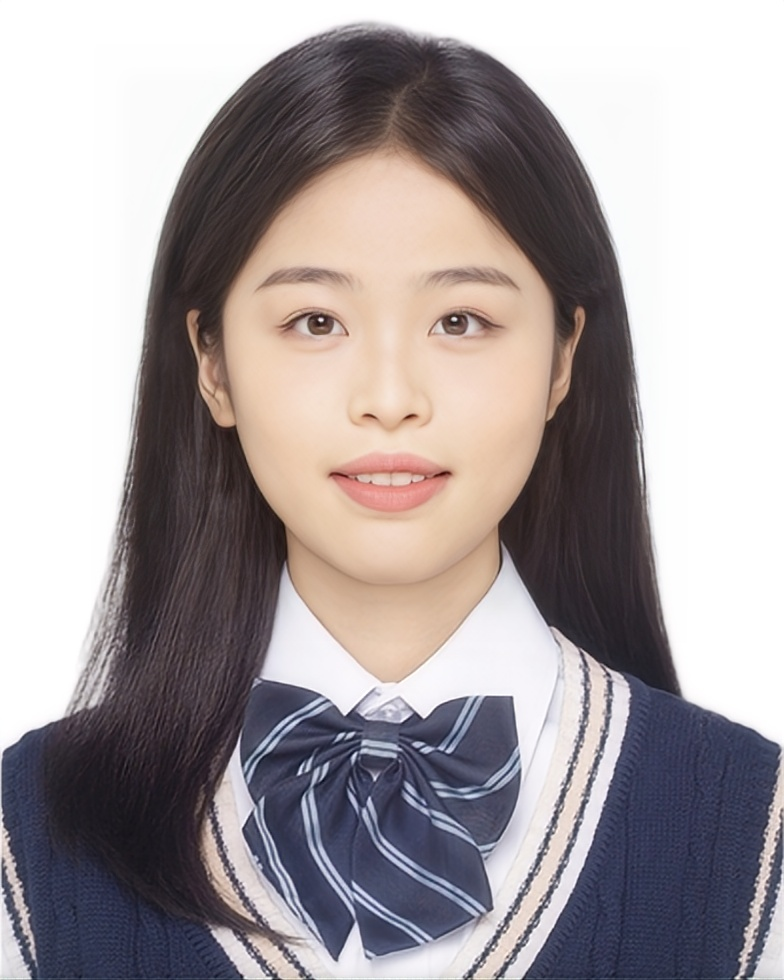}}]{Jiaqi Zhu}
is currently pursuing the B.S. degree in Microelectronics Science and Engineering at the College of Electronic Science and Technology, National University of Defense Technology (NUDT).

Her current research interests include deep learning, radar automatic target recognition, and integrated circuits design.
\end{IEEEbiography}
\vspace{-3em}
\begin{IEEEbiography}[{\includegraphics[width=1in,height=1.25in,clip,keepaspectratio]{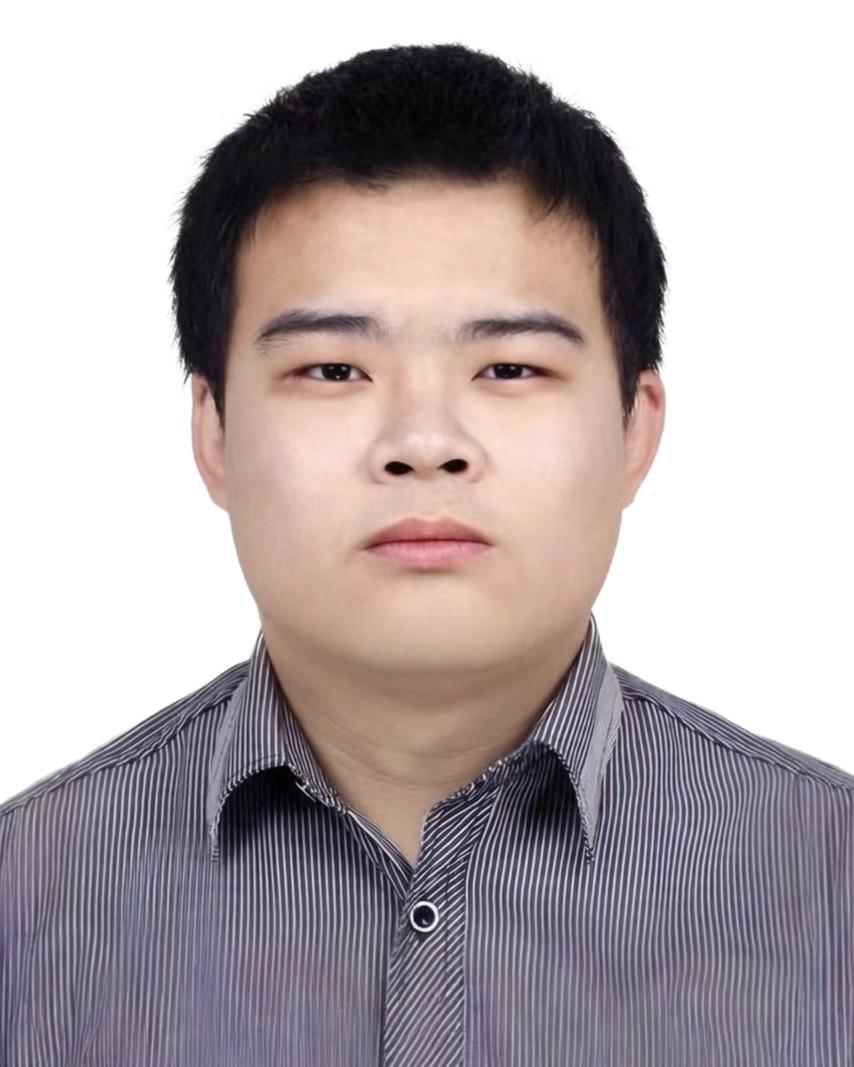}}]{Fengrong Guo}
received the B.S. degree in Electronic Science and Technology from the National University of Defense Technology (NUDT) in 2025.

He is currently pursuing the Ph.D. degree at the College of Electronic Science, NUDT. His current research interests include deep learning and multimodal image fusion.
\end{IEEEbiography}
\vspace{-3em}
\begin{IEEEbiography}[{\includegraphics[width=1in,height=1.25in,clip,keepaspectratio]{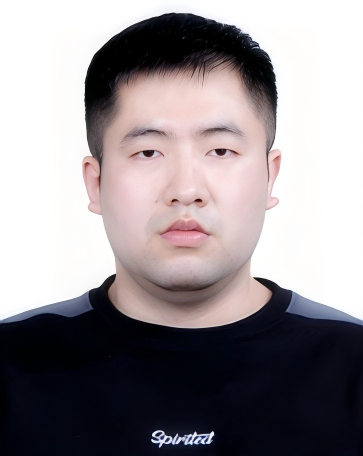}}]{Jiehua Zhang}
received the M.S. degree in aerospace science and technology from the National University of Defense Technology (NUDT), Changsha, China, in 2020, and the Ph.D. degree in computer science and engineering from the University of Oulu, Oulu, Finland, in 2025.

He is currently a Post-Doctoral Researcher with NUDT. His research interests include deep learning, computer vision, and efficient network design.
\end{IEEEbiography}
\vspace{-3em}
\begin{IEEEbiography}[{\includegraphics[width=1in,height=1.25in,clip,keepaspectratio]{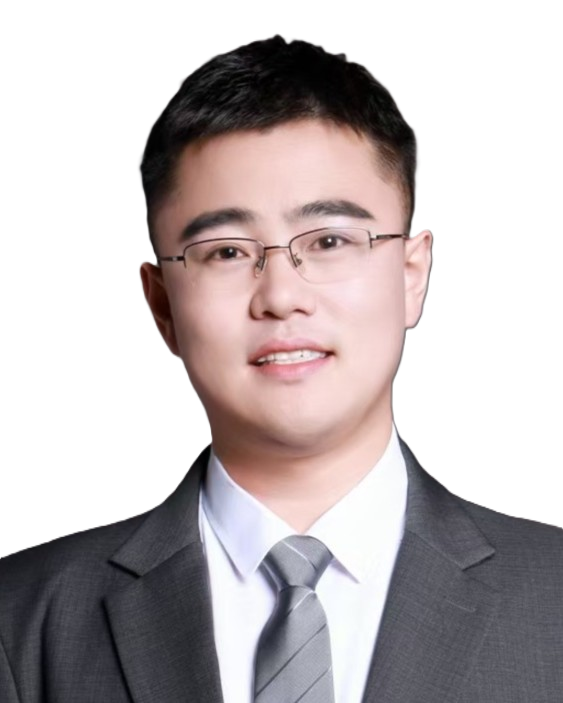}}]{Panhe Hu}
(Member, IEEE) received the B.S. degree in electronic information engineering from Xidian University, Xi'an, China, in 2013 and the Ph.D. degree in information and communication engineering from the National University of Defense Technology (NUDT), Changsha, Hunan, China, in 2019.

He is currently an Associate Professor with the College of Electronic Science and Technology, NUDT. His current research interests include radar system design and array signal processing.
\end{IEEEbiography}
\vspace{-3em}
\begin{IEEEbiography}[{\includegraphics[width=1in,height=1.25in,clip,keepaspectratio]{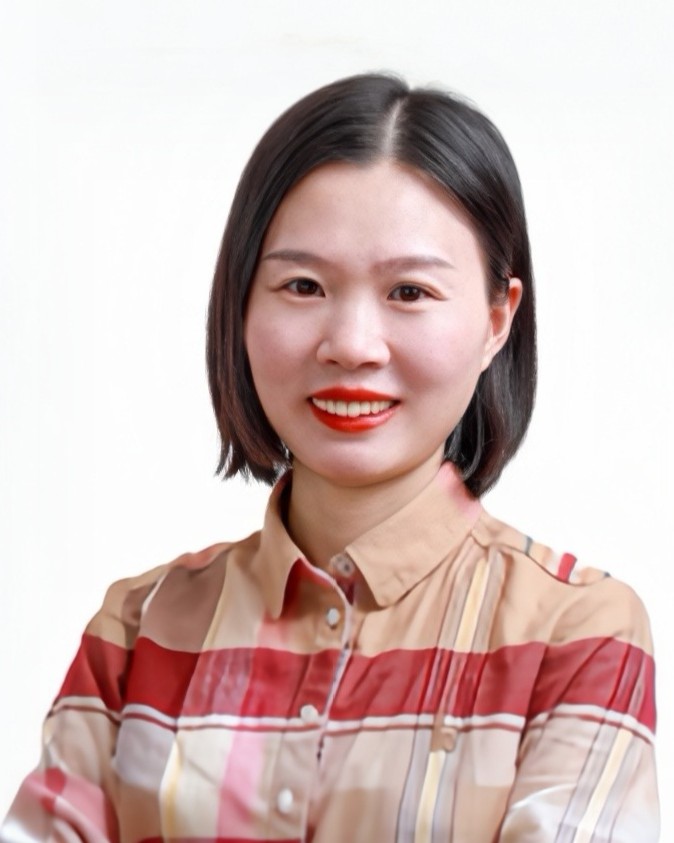}}]{Li Liu}
(Senior Member, IEEE) received the Ph.D. degree from the National University of Defense Technology (NUDT), Changsha, China, in 2012. She is currently a Full Professor with NUDT. She has held visiting appointments with the University of Waterloo, Waterloo, ON, Canada; The Chinese University of Hong Kong, Hong Kong; and the University of Oulu, Oulu, Finland. Her papers currently have over 16,900 citations, according to Google Scholar. Her research interests include computer vision, pattern recognition, and machine learning.

Dr. Liu served as a Co-Chair of many International Workshops along with major venues like CVPR and ICCV. She also served as an Area Chair for several respected international conferences. She served as the Leading Guest Editor of the special issues for IEEE Transactions on Pattern Analysis and Machine Intelligence and the International Journal of Computer Vision. She currently serves as an Associate Editor for IEEE Transactions on Circuits and Systems for Video Technology and \textit{Pattern Recognition}.
\end{IEEEbiography}
\vspace{-3em}
\begin{IEEEbiography}[{\includegraphics[width=1in,height=1.25in,clip,keepaspectratio]{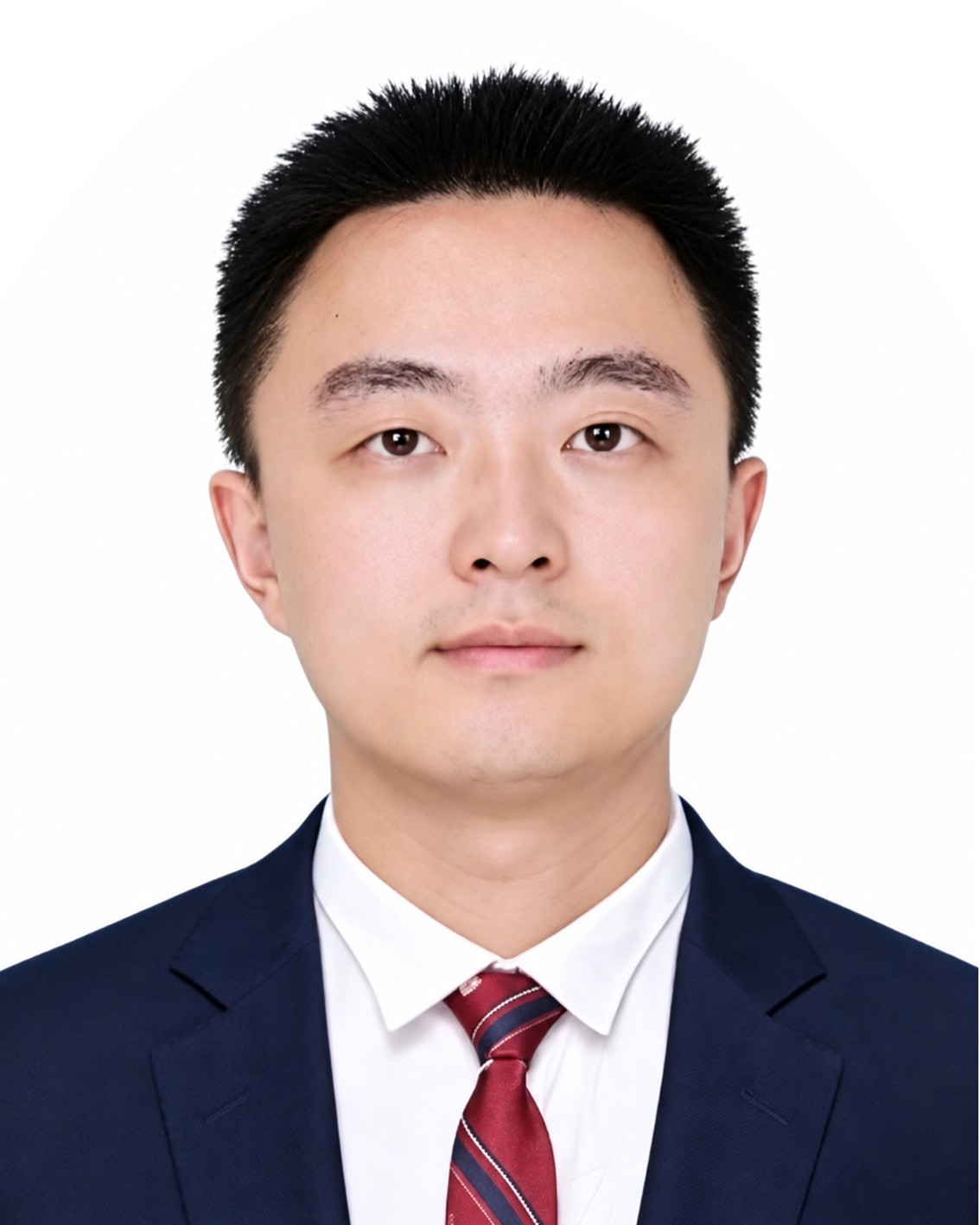}}]{Zhen Liu}
(Member, IEEE) received the M.S., and Ph.D. degree in information and communication engineering from the National University of Defense Technology (NUDT), in 2008 and 2013 respectively. Before that, he received the B.S. degree from Zhejiang University in 2006.

He is currently a professor with the College of Electronic Science and Technology, NUDT. He has been awarded the Excellent Young Scientists Fund on his project titled ``Intelligent Countermeasure for Radar Target Recognition'' in 2020. His current research interests include radar signal processing, radar electronic countermeasure, compressed sensing, and machine learning.
\end{IEEEbiography}
\vfill

\end{document}